\documentclass[a4paper,11pt]{article}
\pdfoutput=1 % if your are submitting a pdflatex (i.e. if you have
\usepackage{jheppub} % for details on the use of the package, please
\usepackage{color}
\usepackage{graphicx}
\usepackage{amsmath}
\usepackage{amssymb}
\usepackage[normalem]{ulem}
\definecolor{darkred}{rgb}{0.6,0,0}

\renewcommand\thesection{\arabic{section}}

\definecolor{linkcolor}{rgb}{0,0,0.5}

\def\gsim{\raise0.3ex\hbox{$\;>$\kern-0.75em\raise-1.1ex\hbox{$\sim\;$}}}
\def\lsim{\raise0.3ex\hbox{$\;<$\kern-0.75em\raise-1.1ex\hbox{$\sim\;$}}}

\def\beqn#1{\begin{equation}\label{#1}}
\def\eeqn{\end{equation}}

\def\beqa#1{\begin{eqnarray}\label{#1}}
\def\eeqa{\end{eqnarray}}

\def\Z2{$\mathcal{Z_2}$}

\def\vev#1{\left\langle #1\right\rangle}

\newcommand {\ignore}[1]{}

\begin{document}
\title{The Stochastic Axiverse}
\author[a]{Mario Reig}\emailAdd{mario.reig@ific.uv.es}
\affiliation[a]{Instituto de F\'isica Corpuscular (CSIC-Universitat de Val\`encia),
	C/ Catedr\'atico Jos\'e Beltr\'an 2, E-46980 Paterna (Valencia), Spain}

%%%%%%%%%%%%%%%%%%%%%%%%%%%%%%%%%%%%%%%%%%%%%%%%%%%%%%%%%

%\begin{textblock*}{5cm}(11cm,-8.2cm)
%\end{textblock*}

%%%%%%%%%%%%%%%%%%%%%%%%%%%%%%%%%%%%%%%%%%%%%%%%%%
%
\abstract{In addition to spectacular signatures such as black hole superradiance and the rotation of CMB polarization, the plenitude of axions appearing in the string axiverse may have potentially dangerous implications. An example is the cosmological overproduction of relic axions and moduli by the misalignment mechanism, more pronounced in regions where the signals mentioned above may be observable, that is for large axion decay constant. In this work, we study the minimal requirements to soften this problem and show that the fundamental requirement is a long period of low-scale inflation. However, in this case, if the inflationary Hubble scale is lower than around $O(100)$ eV, no relic DM axion is produced in the early Universe. Cosmological production of some axions may be \textit{activated}, via the misalignment mechanism, if their potential minimum changes between inflation and today. As a particular example, we study in detail how the maximal-misalignment mechanism dilutes the effect of dangerous axions and allows the production of axion DM in a controlled way. In this case, the potential of the axion that realises the mechanism shifts by a factor $\Delta\theta=\pi$ between the inflationary epoch and today, and the axion starts to oscillate from the top of its potential. We also show that axions with masses $m_a\sim O(1-100)\, H_0$ realising the maximal-misalignment mechanism generically behave as dark energy with a decay constant that can take values well below the Planck scale, avoiding problems associated to super-Planckian scales. Finally, we briefly study the basic phenomenological implications of the mechanism and comment on the compatibility of this type of \textit{maximally-misaligned quintessence} with the swampland criteria.}

\maketitle
\flushbottom

%%%%%%%%%%%%%%%%%%%%%%%%%%%%%%%%%%%%%%%%%%%%%%%%%%%%%%%%%
\newpage

\section{Introduction and summary}
Axions are among the best motivated and most ubiquitous particles beyond the Standard Model. In addition to the QCD axion \cite{,Wilczek:1977pj,Weinberg:1977ma}\footnote{For recent reviews on the QCD axion, see~\cite{diCortona:2015ldu,Irastorza:2018dyq,Hook:2018dlk,DiLuzio:2020wdo} and references therein.}, appearing in the Peccei-Quinn (PQ) solution to the strong CP problem \cite{Peccei:1977hh}, and other models with approximate shift symmetries, a plethora of axions appears in most string compactifications \cite{Arvanitaki:2009fg,Svrcek:2006yi}. With potentially several axions populating every decade of energy, the axiverse has an extremely rich phenomenology. The details of their phenomenological implications mainly depend on the axion mass and the coupling to the visible sector \cite{Arvanitaki:2009fg}. 

In recent years, axion detection has received a lot of attention with a huge experimental effort looking for axions through different portals and at many different mass scales, spanning more than 30 orders of magnitude. Examples include the axion coupling to nucleons and fermions \cite{Budker:2013hfa,Arvanitaki:2014dfa} and the promising avenue to discover axions through their coupling to photons \cite{Wilczek:1987mv,Sikivie:1983ip,Dreyling-Eschweiler:2014mxa,TheMADMAXWorkingGroup:2016hpc,Andriamonje:2007ew,Anastassopoulos:2017ftl,Asztalos:2009yp,Armengaud:2014gea,Majorovits:2016yvk,Kahn:2016aff,Arvanitaki:2017nhi,Baryakhtar:2018doz,Chaudhuri:2018rqn}. While most of these searches depend on some cosmological abundance of relic axions, there exist other strategies that do not depend on their contribution to the energy density budget of the Universe.  For a review on experimental searches see \cite{Irastorza:2013dav,Graham:2015ouw}. It might happen that most of the axions in the axiverse only interact with the visible world through gravitational interactions. 
However,  in string theory, it is generically expected that at least one axion is coupled to QED and QCD through a dimension-5 operator, $\frac{a}{F_a}F\tilde{F}$ and the equivalent gluon operator\footnote{Note, however, that this does not guarantee the axion solution to the strong CP problem in string theory. The validity of the mechanism depends on model building details such as the moduli stabilization mechanism or the presence of string instantons that might spoil the PQ solution \cite{Arvanitaki:2009fg}.}. The reason being that continuous parameters (such as gauge couplings) are determined by the VEV of scalar fields, the moduli fields. The pseudoscalar partner of the modulus that fixes the gauge coupling generically behaves as an axion. An immediate example is the well known model-independent QCD axion of string theories \cite{Fox:2004kb}. On the other hand it may happen that, in addition to the QCD axion, several additional axions in the axiverse are coupled to photons. This motivates to look for axions, which may compose DM or not, beyond the QCD band (see \cite{Agrawal:2021dbo} for a recent review). An example are hyperlight axions with mass $10^{-33}\text{ eV}\lesssim m_a\lesssim 10^{-28}$ eV that generate a rotation of the CMB polarization if they are coupled to QED \cite{Arvanitaki:2009fg,Lue:1998mq,Pospelov:2008gg}. If the inflationary scale is sufficiently large, the associated hyperlight axion strings can also cause a similar effect \cite{Agrawal:2019lkr}, with a rotation angle that is quantized and directly related to the anomaly coefficient.

While most of the phenomenology depends on the axion couplings to the visible sector, some other implications depend only on their gravitational interactions. These string theoretic axions appearing in the axiverse inherit most of their properties directly from the geometric and topological properties of the compactification manifold. Therefore, if sufficiently model independent, bounds on axions can be used to exclude string compactifications. This philosophy has been recently exploited in \cite{Mehta:2020kwu,Mehta:2021pwf}, using the phenomenon of black hole (BH) superradiance  to constrain the landscape of possible theories. BH superradiance is a compelling tool to test the axiverse because it is independent of the axion abundance or other details of the cosmological evolution \cite{Arvanitaki:2009fg,Arvanitaki:2010sy}.
The idea is that the superradiant instability of Kerr black holes efficiently extracts angular momentum from the BH, populating an axion cloud around it. Measuring the properties of fast-spinning BH allows to constrain axion masses with Compton wavelength of the order of the size of the BH, $R_{BH}\sim m_a^{-1}$. However, these constraints become weaker for axion decay constants below $F_a\lesssim10^{13-16}$ GeV, depending on the axion mass, where the self-interaction effect becomes relevant and can suppress BH superradiance spin-down (see \cite{Arvanitaki:2014wva,Baryakhtar:2020gao,Mehta:2021pwf} for detailed, recent studies). 

Another renowned phenomenon that mainly depends on their gravitational interactions is the contribution of stable axions to the dark matter energy density\footnote{Of course, details of the cosmological model or even their couplings to the visible sector may slightly modify their relic abundance.}. Light axions with masses $m_a>10^{-22}$ eV are generically stable and can contribute to the observed dark matter abundance. There exist several mechanisms to produce non-relativistic axions efficiently. For example, if PQ (or equivalent symmetry for generic axions) symmetry is thermally restored after inflation, DM axions can be efficiently produced from the decay of different topological defects \cite{Davis:1986xc,Vilenkin:2000jqa}. See also \cite{Klaer:2017ond,Gorghetto:2018myk,Buschmann:2019icd,Gorghetto:2020qws} for recent simulations of the axion string contribution. Instead, when PQ is never restored after reheating, the misalignment mechanism \cite{Preskill:1982cy,Abbott:1982af,Dine:1982ah}, i.e. coherent oscillations around the minimum of the potential, is the dominant contribution. For a generic axion appearing in the axiverse\footnote{From now on we will assume an axion potential of the type: $V(a)=m_a^2F_a^2\left(1-\cos\left(a/F_a\right)\right)$.}, the relic abundance set by these oscillations is approximately given by:
\begin{equation}\label{eq:abundance}
\Omega_a h^2\sim 0.05\,\theta_i^2\left(\frac{F_a}{10^{17}\text{ GeV}}\right)^2\left(\frac{m_a}{10^{-22}\text{ eV}}\right)^{1/2}\,.
\end{equation}
The upshot is that while DM axions with $m_a\ll 10^{-22}$ eV are severely constrained \cite{Hlozek:2014lca}, axions with $m_a\gg 10^{-22}$ eV can overproduce DM depending on their decay constant. The model independent axion of string theory, with a decay constant around $F_a\sim 1.1\times 10^{16}$ GeV, is a prototypical example of well-motivated scenarios where we have axion DM overproduction. Its contribution can be \textit{regulated} by requiring an initial misalignment angle smaller than around $\theta_i\lesssim 10^{-3}$ when PQ is broken before inflation \cite{Fox:2004kb}. The post-inflation case, where one takes the average of misalignment angles over different Hubble patches, is excluded in standard cosmologies due to DM overproduction.

As we have argued above, because of the effect of axion self-interactions, BH superradiance is generically easier to observe in regions of the parameter space where axions are overproduced via the misalignment mechanism. It seems therefore likely that if some signal related to superradiance is observed - as for  example a GW signal from the axion cloud for an axion mass around $m_a\sim 10^{-12}$ eV - it will point to a situation with moderate to large decay constant. This may imply that such axion has to be tuned with $\theta_i\ll 1$ in order to have an acceptable relic production \cite{Arvanitaki:2010sy}. As we will see later, the situation becomes more troublesome as the number of axions increases. The study of situations where there exist a large number of axions overproducing DM - and how to regulate their contribution -, is the main goal of this work. \\\\
On the other hand, the axions that are light enough will have not started to oscillate. In general, for a random initial value for the axion field, this occurs whenever the mass is comparable or smaller than the Hubble scale today, $m_a\lesssim H_0\simeq1.4\times 10^{-33}$ eV. Such axions are still frozen today by Hubble friction and contribute to the energy density budget of the Universe in a way that is almost indistinguishable from a cosmological constant (CC), this is with equation of state parameter $w\equiv p/\rho=-1$. Axions slightly heavier than $H_0$, but still lighter than around $10^{-32}$ eV to prevent premature oscillations, will have started to displace towards the minimum and, in a slow-roll state, contribute to dark energy producing small departures from the CC prediction \cite{Frieman:1995pm} (see also \cite{Copeland:2006wr} for a comprehensive review). The equation of state of these quintessence axions is constrained to be $w \lesssim-0.95$ (95\% C.L.) based on Planck 2018 data \cite{Aghanim:2018eyx}. 

Light scalar fields are generically plagued by fine tuning problems. Unlike their scalar partners, the mass of light axions is protected from large quantum corrections by a shift symmetry, only broken non-perturbatively. However, even in the string axiverse where light axions that can play the role of dark energy are abundant \cite{Svrcek:2006hf}, successful axion quintessence has to overcome a number of challenges \cite{Cicoli:2018kdo}. The most obvious is that the minimum of the axion potential describing dark energy has to be tuned so that its minimum is at around zero energy value while its maximum has a height of $V(a)_{max}\sim \rho_{vac}=(3\text{ meV})^4$. 
Also, it has  been pointed out that axion quintessence generically needs $F_a> O(1)M_P$. The reason for requiring this large decay constant is twofold. If the initial axion field value lies in the convex region of the axion potential, $F_a> O(1)M_P$ is required to satisfy the slow-roll condition. On the other hand, starting in the concave region requires a decay constant comparable or larger than $M_P$ to overcome the tachyonic instability of the axion potential \cite{Kaloper:2005aj}. The tachyonic instability can be compensated by requiring that the initial axion field value is extremely close to the top of the potential, $a/F_a\approx \pi$, where the first derivative nearly vanishes. As explained in \cite{Kaloper:2005aj}, this requires an exponential tuning of the initial misalignment angle. We will show that in certain situations this initial condition can be dynamically achieved.

As it is well-known in the context of theories of axion inflation, the requirement of transplanckian decay constants generically brings associated problems in the effective theory. This occurs, for example, due to the presence of low-action instantons that spoil the flatness of the potential by introducing unsuppressed higher harmonics \cite{Banks:2003sx}.
Additionally, with the advent of the string swampland criteria (see \cite{Palti:2019pca} and references therein), it is currently unclear if the standard picture of the Planck-size decay constant required to fight the tachyonic instability of the axion potential close to its maximum is compatible with string theory as a theory of quantum gravity. Despite it may seem that the collective effect of many axions with $F_a\ll M_P$ is consistent with the criteria, a change of basis reveals that in such theories, transplanckian scales are hidden in a basis rotation \cite{Rudelius:2015xta,Montero:2015ofa,Heidenreich:2015wga}. As an example, theories such as N-flation \cite{Dimopoulos:2005ac}, qui$\mathcal{N}$tessence \cite{Kaloper:2005aj} and other multi-axion models are in conflict with the distance conjecture when the traversed distance exceeds $\sim O(1)$ in Planck units or, in other words,  $\sqrt{N}F_a>O(1)\,M_P$ \cite{Agrawal:2018own}. 
\subsection{Summary}
In this work we tackle both issues - that is, the DM axion overabundance and the apparent difficulty in finding consistent axion quintessence -, together in the axiverse.  First of all, in section \ref{sec_stochastic_axiverse}, to illustrate the overabundance problem we study a simplified picture of the axiverse with \textit{geometric} axions. These axions have masses and decay constants determined exclusively by geometric reasons, allowing to understand in a transparent way why their abundance is troublesome in compactifications that realise gauge sectors with couplings comparable to the SM gauge couplings. After that, we show that a long period of low-scale inflation suffices to \textit{regulate} the contribution to the energy density of these axions. In a similar way to \cite{Graham:2018jyp,Guth:2018hsa}, given an inflationary Hubble scale $H_I$, axions with potentials satisfying $H_I\lesssim (m_aF_a)^{1/2}$ will slow-roll to their minimum during inflation and reach an equilibrium configuration where the inflationary quantum fluctuations generate some distribution around the minimum of the potential. Despite their contribution to the energy density of the Universe is not exactly zero, it may be arbitrarily suppressed if the inflationary Hubble scale is sufficiently low. By analogy, we call this scenario \textit{The Stochastic Axiverse}.

We will show in section \ref{sec_stochastic_axiverse} that if the inflationary Hubble scale is smaller than a reference value that lies around $H_I\sim O(100)$ eV, no relevant abundance will be ever produced by the misalignment mechanism. The right abundance of axion DM can be recovered if the potential minimum of a given axion (or at most few of them) changes between inflation and today. To illustrate this effect, in section \ref{sec:toy_model} we generalise the mechanism presented in \cite{Huang:2020etx} that allows an initial misalignment angle close to maximal. In section \ref{sec:applications} we will apply the mechanism to axions in different mass regimes. For axions with masses $m_a>10^{-22}$ eV, this mechanism allows us to have axion DM with all the phenomenology associated to the large-misalignment scenario studied in \cite{Arvanitaki:2019rax}. When applied to axions with masses comparable to the Hubble scale today, $m_a\sim O(1-100)H_0$, we show that the same mechanism allows us to obtain axion quintessence with a decay constant substantially below the Planck scale and, in some cases, close to the GUT scale. The reason, as we will see later, is that the mechanisms presented in section \ref{sec:toy_model} allow to have an initial axion field value arbitrarily close to the top of its potential, compensating the tachyonic instability of axion potentials with $F_a\ll M_P$. 

In section \ref{sec:pheno} we study the basic phenomenology associated to the mechanism of maximal axion misalignment, where gauge sectors with low confinement scales are present. We will consider the effects that these new, dark gauge sectors have on the effective number of neutrinos, $\Delta N_{eff}$, and why isocurvature fluctuations actually constrain the mass of the axion DM candidate even if the inflationary scale is very low. Later, in section \ref{sec:string_model_building}, we briefly study the mechanism in a string theory model building context showing how the requirement of low inflationary Hubble scales, and therefore relatively low reheating temperatures, offers indications about the string scale and the axion couplings to the visible gauge sectors.  Finally, in section \ref{sec:Discussion} we discuss the possibility of alternative ways to regulate the axiverse and comment on the compatibility with the string swampland criteria. 
\begin{figure*}[t]
	\centering
	\includegraphics[width=0.69\textwidth]{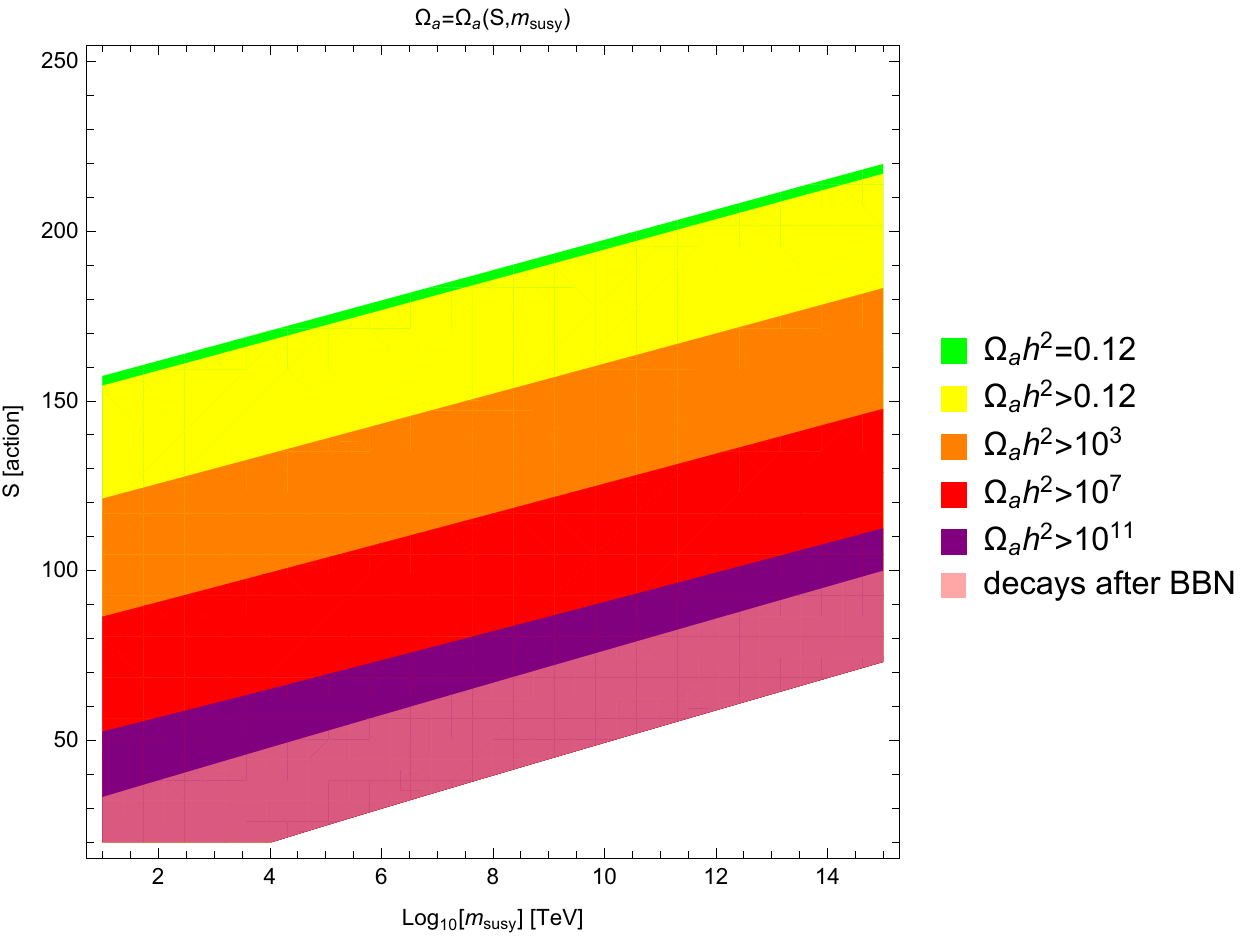}
	\caption{Relic abundance as a function of the action $S$ and the SUSY breaking scale $m_{susy}$ for a single geometric axion. Instanton actions below $S\sim 150$ tend to overproduce axions and, depending on the SUSY breaking scale, the required tuning can be severe. The tuning in the initial condition to compensate the overabundance goes as $\theta_{tun}\sim 1/\sqrt{\Omega_a}$. As expected, each axion suffers of this problem individually, resulting requiring a huge tuning if the number of axions is $n> O(10)$. In the region above the thin, green band axions compose a subdominant part of the DM. Assuming a decay contant around the GUT scale, $F_a\sim M_{GUT}$, in the shaded pink region the axion decays between BBN and today, spoiling the success of nucleosynthesis. Below this region, axions decay before BBN and are in principle cosmologically allowed.}
	\label{fig_relic_density}
\end{figure*}
\section{The Stochastic Axiverse}
\label{sec_stochastic_axiverse}
Let us now systematically analyze a simplified picture of the axiverse. In string theory constructions, gauge couplings are determined by the VEV of moduli fields. For example, if one takes a type-II theory, the gauge coupling is inversely proportional to the volume of the cycle in string units, $\tau$, that the D-brane wraps:
\begin{equation}
\frac{1}{\alpha}=\tau\,.
\end{equation}
If we want to realise the SM, we need some distribution of the couplings around the standard GUT value, $\alpha_{GUT}\sim 1/25$, at the compactification scale. As argued in \cite{Svrcek:2006yi,Banks:2003sx}, many of the axions will have a potential and a decay constant determined exclusively by geometric reasons:
\begin{equation}\label{geometrical_axion_parameter}
V(a)\sim \Lambda^4e^{-S}\cos\left(a/F_a\right)\,,\,\,\,\, F_a\sim M_P/S\,.
\end{equation}
Here $S\sim 2\pi/\alpha$ is the action of the string instanton or dark Yang-Mills (YM) sector that dominates the axion mass and $\Lambda^4\sim m_{susy}^2M_P^2$, with $m_{susy}$ the scale of SUSY breaking. These \textit{geometric} axions saturate the bound of the Weak Gravity Conjecture \cite{ArkaniHamed:2006dz} and, while it is true that the decay constant can be reduced in warped or highly anisotropic compactifications, as we will see now, they are a good guide to illustrate the issue of overabundance.
	
From the quantities above, one recovers the typical expectation that axion masses have an approximately flat distribution in a log scale while the decay constant obeys an approximately normal distribution around a typical value. We also see that, while obtaining almost massless axions (those with masses $m_a\lesssim H_0$) is relatively simple because of the exponential dependence on $\alpha^{-1}$, it is challenging to obtain axion decay constants below $M_{GUT}$ unless the axion is coupled to an hyperweak dark sector with $\alpha\ll 10^{-2}$. Therefore, typical decay constants ranging from $M_{GUT}\lesssim F_a\lesssim M_P$ are expected to arise in the axiverse in compactifications where cycles that can realise the SM are not rare. 
		
It has been argued that at least several string axions may behave as fuzzy DM \cite{Hui:2016ltb}. 
The relic abundance of these typical string axions behaves (assuming a mild temperature dependence of the axion potential) as $\Omega_a\propto F_a^2 m_a^{1/2}$. 
The abundance gets enhanced if the axion potential has a strong temperature dependence, as the QCD axion, being very suppressed at high temperatures. This occurs because the temperature suppression of the potential slightly delays the onset of oscillations. If, for simplicity, we consider the case of the geometric axions introduced in Eq. (\ref{geometrical_axion_parameter}) one can write the relic abundance of the axion as a function the action and the SUSY breaking scale:
\begin{equation}
\Omega_a h^2\sim 0.05\,\theta_i^2\left(\frac{200}{S}\right)^{3/2}\left(\frac{m_{susy}}{10\text{ TeV}}\right)^{1/2}\exp\left[\frac{155-S}{4}\right]\,.
\end{equation}
The expression above seems helpful to understand the overabundance problem in a more transparent way. Once the SUSY breaking scale - which is a universal parameter, i.e. common for all the axions - is fixed, the relic density of any geometric axion is just a function of a single parameter: the action $S$ of the associated instanton. Alternatively one might visualize this as a function of the volume of the cycles or, equivalently, as a function of the gauge coupling at the compactification scale, since we consider instantons with action, $S\sim 2\pi/\alpha$. 
%We see that, as expected, the relic density is exponentially sensitive to the action: 
%
%\begin{equation}
%\Omega_a\propto m_{susy}^{1/2}S^{-3/2}e^{-S/4}\,,
%\end{equation}
%
%and also increases with the square root of the SUSY breaking scale. 
As an example, we easily see that axions with $S\lesssim 160$, for $m_{susy}\sim 10$ TeV, tend to overproduce DM. Given that the action corresponding to the GUT gauge coupling is $S_{GUT}\sim \frac{2\pi}{\alpha_{GUT}}\sim 155$, it seems plausible that in scenarios with $O(100)$ axions at least a handful of axions will have $S\lesssim 150$, with the subsequent problem of axion overproduction. 
The situation unfortunately gets worse for high-scale SUSY breaking as expected in gravity mediated scenarios, where $m_{susy}$ is expected to be much larger than $O(10)$ TeV. In Fig. \ref{fig_relic_density} we plot the relic density as a function of $S$ and $m_{susy}$ under the assumptions in Eq. \ref{geometrical_axion_parameter}, that is for a single \textit{geometric} axion.

The issue of overabundance is exacerbated by the presence of additional light scalar moduli \cite{Banks:1993en,deCarlos:1993wie} (see \cite{Fan:2011ua} for a more recent study). In some mechanisms of moduli stabilisation (non-SUSY mechanisms) light axions usually come with associated light scalar moduli. The mass of the lightest moduli typically depends on the SUSY breaking scale, $m_\phi\sim m_{susy}^2/M_P$, and have Planck suppressed couplings to matter \cite{Giudice:1998bp}. Being very efficiently produced from coherent oscillations around the minimum, scalar moduli can be dangerous if their life-time is longer than the age of the Universe. 
If SUSY breaking occurs at a high scale, as it is the case for gravity mediated breaking, then their mass is expected to be at the weak or even TeV scale. Of course, these moduli will eventually decay. The problem in this case is that they can decay during or after BBN spoiling the success of nucleosynthesis. In \cite{Arvanitaki:2009fg} the authors consider that a Hubble scale of $H_I\lesssim 0.1$ GeV allows the moduli to go to the minimum during inflation so that they are never produced (thermally and non-thermally). This might not be enough in gauge mediated SUSY breaking, where the light moduli typically have masses between $10 - 10^6$ eV \cite{Fan:2011ua}. These moduli are expected to be cosmologically stable and will overclose the Universe unless their abundance is depleted. We will show that the stochastic axiverse, with a substantially lower $H_I$, automatically regulates these moduli.

There are in principle few ways out to this problem. It might happen that the number of axions is $O(1)$ and, for some reason, they lie 
%in the sweet-spot of $F_a\sim 10^{17}$ GeV and $m_a\sim 10^{-22}$ eV. This is in some sense equivalent to demanding that our axions fall precisely 
in the thin green band in Fig.\ref{fig_relic_density}. On the other hand, when the number of axions is $O(1)$, one can evoke the anthropic principle to explain why the axion abundance matches the observed DM abundance \cite{Tegmark:2005dy,Freivogel:2008qc}. In this case, for example, anthropics would accommodate the small initial misalignment angle that allows a QCD axion with a decay constant $F_a\sim M_{GUT}$.  
However, while anthropic arguments may help when the number of axions is small, the necessity for tuning becomes unavoidable when the number of axions is $n>O(10)$. For example, using the same measure as in \cite{Arvanitaki:2010sy}, a theory with $n\sim 30$ axions with $F_a\sim 10^{16}$ GeV and $m_a\geq 10^{-19}$ eV needs a fine tuning at the level of $\sim 10^{-20}$ after applying the anthropic argument. Despite the concrete number has some dependence on the choice of measure, as explained in \cite{Arvanitaki:2010sy} the fact that for a large number of axions the tuning grows exponentially fast seems to be measure independent.

Clearly the anthropic explanation fails with a large number of axions, motivating a dynamical solution. In this direction, it could happen that all the decay constants are small ($F_a\ll M_{GUT}$) due to highly warped or anisotropic compactification manifolds. When applied to a large number of axions, this solution requires that only the SM cycle, and at most few others, have a volume that is not extraordinarily large in string units. This would render our visible sector a rare region of the compactification manifold. 
Finally, there is of course the possibility that the initial misalignment is small, $\theta_i\ll 1$, for some dynamical reason. In the rest of the paper we will consider this possibility, which appears to be rather independent of the details of the topology of the compact manifold. We will study the minimal requirements of this kind of solutions and, later, study in detail a particular model based on the maximal-misalignment mechanism presented in \cite{Huang:2020etx}. 
\subsection{Dynamical ways out}
As we anticipated, we can regulate the contribution of axions to the energy density by imposing $\theta_i\ll 1$. While for a $O(1)$ axions this assumption is not completely unreasonable, the amount of tuning we need grows very fast as we increase the  number of axions. We are therefore challenged to find a dynamical answer to this question. Being a problem related to initial conditions, possible answers will be tightly connected to the inflationary epoch. In particular, we consider theories where the axion field relaxes to the minimum of its potential during inflation. Two possibilities arise. The first one is that the inflationary axion masses are tremendously larger than their current value so that these axions start oscillating during inflation and eventually relax to their minimum \cite{Dvali:1995ce}. This scenario in principle suffers from a number of difficulties \cite{Choi:1996fs} that we will discuss in section \ref{sec:Discussion}. 

The second possibility, recently proposed in \cite{Graham:2018jyp,Guth:2018hsa} in the context of the QCD axion, is that the inflationary Hubble scale is smaller than the current height of the axion potential:
\begin{equation}\label{Hubble_condition}
H^4_I < (m_aF_a)^2\,,
\end{equation}
so that the potential is turned on during inflation and, even though $m_a \ll H_I$, the axion slowly rolls towards the minimum \cite{Dimopoulos:1988pw}. In this case, since the Hubble parameter during inflation is smaller than the height of the vacuum axion potential, the axion potential is turned on during inflation and both minima - the inflationary and today's axion vacuum - are expected to coincide. In recent years, this stochastic axion scenario has received some attention in different contexts \cite{Ho:2019ayl,Marsh:2019bjr,Takahashi:2019qmh,Nakagawa:2020eeg}. 
%In particular, the authors of \cite{Graham:2018jyp,Guth:2018hsa} considered a similar scenario with a long and low-scale inflationary period as a mechanism to allow a QCD axion with a large decay constant - up to the Planck scale -, removing the upper bound of the QCD axion window. 
In \cite{Ho:2019ayl}, partially overlapping with this work, a similar mechanism of long and low-scale inflation was applied to string axions with decay constant around $10^{16}\text{ GeV}$ and masses $10^{-18}\text{ eV}\lesssim m_a\lesssim 10$ TeV. Before applying the  mechanism to the many axions of the axiverse let us briefly overview the basic inflationary dynamics for a single axion case. 

In order to relax down to the minimum of the axion potential, the inflationary Hubble parameter has to satisfy $H_I<(m_aF_a)^{1/2}\sim\Lambda_D$, so that the potential is turned on. In this case, the axion field slowly rolls towards the minimum. The equation of motion (EOM) is given by:
\begin{equation}
\ddot{a}+3H\dot{a}+V'(a)=\eta(t)\,,
\end{equation}
with $\eta(t)$ the stochastic noise due to inflationary quantum fluctuations. The axion experiments two effects that balance with each other \cite{Linde:2005ht}: 
\begin{itemize}
	\item \textbf{Slow-roll}: the axion is very slowly rolling towards the minimum with EOM: 
	\begin{equation}
	\dot{a}=-\frac{V'(a)}{3H_I}\rightarrow a(t)\propto e^{-m^2 t/3 H_I}\,.
	\end{equation}
	\item \textbf{Diffusion}: Quantum fluctuations kick the axion field, jumping randomly by a factor $H_I/2\pi$ in a Hubble time. 
\end{itemize}
After a long inflationary period ($N_e> H_I^2/m_a^2$), the combination of slow-roll and diffusion/random walk of the axion field arrive to an equilibrium configuration \cite{Graham:2018jyp,Guth:2018hsa}. For the cosine potentials we are considering throughout the paper, $V(a)\propto \left(1-\cos\left(a/F_a\right)\right)$, the mean is at zero field value, $\vev{a}=0$, with a \textit{spread} given by:
\begin{equation}\label{small_misalignment}
\frac{\delta a}{F_a}= \delta \theta = \sqrt{\frac{3}{8\pi^2}}\frac{H_I^2}{m_aF_a}\,.
\end{equation}
Notice also that due to inflation, the initial condition for the axion field is homogeneous in all our Universe. In typical models that we will consider later, the main source of shift symmetry breaking will be the instantons of a confining, dark YM interaction, resulting in: $m_aF_a\sim \Lambda_D^2$. Depending on the confinement scale of the dark non-abelian group, $\Lambda_D$, the phenomenology and implications for cosmology are intriguing and diverse. 
In particular we focus on the limit of ultralight axion masses which potentially have interesting implications for DM (for $m_a\sim 10^{-22}$ eV and larger) and the accelerated expansion of our Universe, when $m_a$ is comparable to $H_0$ \cite{Arvanitaki:2009fg,Svrcek:2006yi}. 

To illustrate this mechanism of low-scale inflation let us consider a light axion that contributes to the relic density of DM. In the standard misalignment case, one expects an initial angle $\theta_i\sim O(1)$. The observed relic density, $\Omega_a\approx\Omega_{DM}=0.12$, is then obtained for a decay constant that we define as $F_a^{mis}$. The concrete value will of course depend on the axion mass, $m_a$ and other details of the potential such as the temperature dependence. As explained in \cite{Graham:2018jyp,Guth:2018hsa}, for any given axion, once we take into account the long and low-scale inflationary period, the situation can be classified in three different regimes. 

For a large decay constant:
\begin{equation}
F_a\gg F_a^{mis}\,,
\end{equation}
it may happen that $H_I$ is moderately large (but still satisfying the condition \ref{Hubble_condition}), leading to $\theta_i\lesssim O(0.1-1)$, and no strong suppression of the relic abundance is achieved. This situation resembles the standard misalignment scenario with a large decay constant and will require additional tuning to beat the sizeable spread of the wave function around the minimum generated by moderately large quantum fluctuations. For a single axion in this region, known as the \textit{anthropic axion}, we can accomodate the observed DM abundance by using  anthropic arguments \cite{Tegmark:2005dy,Freivogel:2008qc,Arvanitaki:2010sy}. The opposite possibility is that, for a given decay constant that may be large $F_a\gg F_a^{mis}$, the inflationary Hubble parameter is very low, $H_I\ll (m_aF_a)^{1/2}$, and the axion abundance is extremely suppressed by a factor: 
\begin{equation}
\Omega_a\propto\delta\theta^2\sim H_I^4/(m_aF_a)^2\,.
\end{equation}
This situation leads to an insignificant amount of relic axions. Finally, there is an intermediate region in which, for a given $m_a$ and $F_a$, $H_I$ has the right size to achieve the observed relic abundance. This is known as the \textit{stochastic window} \cite{Graham:2018jyp} and allows a decay constant remarkably larger than the standard misalignment scenario, $F_a\gg F_a^{mis}$, due to the small initial misalignment $\theta_i$ that is achieved after the long period of inflation. 

In the context of the axiverse, where we expect up to several axions per decade of energy from $H_0$ up to $M_P$, this situation may require some amount of tuning of the inflationary Hubble scale. Notice that a too low $H_I$ will completely preclude the existence of axion DM while a too large $H_I$ will reintroduce the problem of overabundance. The reason is that, as we have argued before, anthropic arguments for the initial misalignment angle $\theta_i$ become insufficient in the limit of a large number of axions, and as $H_I$ grows the number of axions with initial random misalignment will increase. The reference value is around $H_I\sim O(100)$ eV and corresponds to an ultralight axion with decay constant $F_a\sim 10^{17}$ GeV and $m_a\sim 10^{-22}$ eV. According to Eq.(\ref{eq:abundance}), for typical decay constants that appear in string compactifications, $M_{GUT}\lesssim F_a\lesssim M_P$, axions with masses such that $\Lambda=(m_aF_a)^{1/2}>100$ eV will typically overproduce DM\footnote{Of course this can be alleviated in the presence of warping of anisotropic manifolds where $F_a$ can be reduced. In such a case, the \textit{reference value} for $H_I$ will slightly grow accordingly.}. Notice also that DM axion masses $m_a\ll 10^{-22}$ eV, are highly disfavored by observations \cite{Hlozek:2014lca}. 

We will describe in the next section how the stochastic axiverse can be minimally modified in a way that is free from the problems above: no axion/moduli overabundance problem without any tuning of $H_I$ nor the use of anthropic arguments. These mechanisms will allow axion DM of any mass as long as $H_I\lesssim O(100)$ eV but without needing any particular value. At the same time, the mechanisms presented here allow a QCD axion with arbitrary decay constant and a superradiant signal at any axion mass.  The basic requirement, as we are about to show, is a long and low-scale inflationary period.

\subsection{Changing the inflationary axion potential}
If all axions settle at their minimum, with small spread of the wave function around it, then the produced DM abundance is negligible. 
Using Eq.(\ref{small_misalignment}), valid for $H_I<(m_aF_a)^{1/2}$ and therefore small $\delta\theta$, we obtain a qualitative expression of the relic density of axions as a function of $H_I$ and $m_a$: 
\begin{equation}\label{relic_small_misalignment}
\Omega_a h^2 \sim 0.05 (\delta\theta)^2 \left(\frac{F_a}{10^{17}\text{ GeV}}\right)^2\left(\frac{m_a}{10^{-22}\text{ eV}}\right)^{1/2}\sim 2\times 10^{-3}\left(\frac{H_I}{10^2\text{ eV}}\right)^4\left(\frac{10^{-22}\text{ eV}}{m_a}\right)^{3/2}\,.
\end{equation}
A remarkable thing has just occured. While the initial relic density depended on 3 parameters that are in principle different for each axion in the axiverse, the small misalignment after a long inflationary period allows to link these parameters (see Eq. (\ref{small_misalignment})). After a bit of algebra, the relic density of each axion does only depend on a \textit{global} parameter that is common to all axions, the inflationary Hubble scale $H_I$, and an axion dependent parameter, the axion mass $m_a$. This result should be taken only as a qualitative indication, but it allows us to see how different axions, with different masses, contribute to the relic density in the stochastic axiverse scenario in a transparent way.
 
This confirms our previous estimate of the reference value for the inflationary Hubble scale. We see that for inflationary Hubble scales below $O(100)$ eV, axions heavier than $m_a\sim 10^{-22}$ eV (the lightest DM axion mass allowed by observations \cite{Hlozek:2014lca}) yield a tiny contribution to the energy density of the Universe. The same mechanism that solves the problem of overabundance of axions and scalar moduli precludes the possibility of having axion DM. However, superradiance and the rotation of CMB polarization may still be observable. This is because BH superradiance does not require any cosmic axion abundance and, also, hyperlight axions will not have relaxed if they are sufficiently light. The latter occurs for light axions with $(m_aF_a)^{1/2}\lesssim H_I$, which have random initial misalignment angles $\theta_i$. 

One way to recover axion DM is to assume that $H_I$ is tuned so that there is an axion (or few of them) with the right mass and decay constant that has not relaxed to its minimum. This could happen if we have an ultralight axion, with $F_a\sim 10^{17}$ GeV and mass $m_a\sim 10^{-22}$ eV, and a Hubble scale during inflation around $H_I\sim 100$ eV. 
According to Eq.(\ref{relic_small_misalignment}), independently of their decay constant, all other axions heavier than around $10^{-22}$ eV will have a negligible impact on the relic density.
This requires to have the right value for the inflationary Hubble scale. If the inflationary Hubble scale is smaller, the ultralight axion abundance fastly decreases as $H_I^4$, while if $H_I\gg O(100)$ eV we reintroduce the problem of overabundance, as more and more axions will have random initial misalignment angles. In some sense, the situation seems to be fine tuned. 
It is tempting to imagine that anthropic selection might explain why $H_I\sim O(100)$ eV so that there is no axion overabundance and only the right amount is produced. The author is not aware of such an anthropic argument for the inflationary Hubble scale $H_I$ in the literature, if possible at all.

Alternatively, one can imagine that all \textit{relevant} axions (from the DM abundance point of view) slow-roll to the minimum due to a small inflationary Hubble scale, $H_I< O(100)$ eV, and then some effect shifts the minimum of a given axion potential today by a quantity $\Delta\theta$ with respect to the inflationary potential. Notice that now, instead of a particular value for $H_I$, we just require that it is bounded from above. In this case, the associated initial misalignment angle for that \textit{activated} axion is given by:
\begin{equation}
\theta_i\approx\Delta\theta\,.
\end{equation}
There are in principle different possibilities to produce this shift. For the sake of completeness, let us consider a simple example where the axion has, during inflation, a two cosine potential:
\begin{equation}
V(a)=-\Lambda_1^4\cos(a/F_a)-\Lambda_{inf}^4\cos(a/F_a-\delta)\,.
\end{equation}
The $\Lambda_{inf}$ part of the potential is generated only during inflation due to some unknown dynamics. Now, imagine we have a situation where $H_{I}\ll\Lambda_{inf}\sim\Lambda_1$. As we remarked above, if $H_I$ is sufficiently small, irrespective of the concrete value, all potentially dangerous axions slow-roll towards the minimum during inflation. After reheating, the part proportional to $\Lambda_{inf}^4$ turns off and an initial misalignment angle has been dynamically generated. Without loss of generality\footnote{If $\Lambda_1<\Lambda_{inf}$, the misalignment reads $\Delta\theta\simeq\frac{\Lambda_1^4}{\Lambda_{inf}^4}\sin\delta$.}, we can assume $\Lambda_1>\Lambda_{inf}$ and this dynamical misalignment will be given by:
\begin{equation}\label{initial_angle_toy_model}
\Delta\theta\simeq \frac{\Lambda_{inf}^4}{\Lambda_1^4}\sin\delta\,.
\end{equation}
We have generated an initial misalignment - that can be potentially small and independent of $H_I$ - in a natural way. In the case $H_I$ is small, for example $H_I\lesssim O(10)$ eV, then only \textit{activated} axions with a mechanism that generates $\Delta\theta\neq 0$ will contribute to DM abundance.

In the rest of the paper we consider the maximal-misalignment mechanism presented in \cite{Huang:2020etx} as a benchmark possibility to turn on the desired axions while shutting down the contribution of potentially dangerous axions. We will study in detail the implications of this concrete mechanism of maximal axion misalignment that generates $\Delta\theta=\pi$ in the context of the axiverse. Incidentally, we find that in addition to axion DM production, the maximally-misaligned axiverse can successfully describe quintessence as dynamical dark energy and we also comment on its compatibility with the WGC and the string swampland criteria.

\section{A toy model of maximal axion misalignment}
\label{sec:toy_model}
Different realisations where the axion starts to oscillate from the top of its potential have been recently proposed \cite{Co:2018mho,Takahashi:2019pqf,Huang:2020etx}. Despite the proposed models look very different, they follow the same philosophy - that is, the axion field relaxes to the inflationary minimum which is later shifted by a factor $\Delta\theta=\pi$ due to some particular dynamics. In this section we review the basics of the mechanism proposed in \cite{Huang:2020etx} for a general, dark gauge group. Consider a non-abelian group $G_D$ with a Dirac fermion $\psi$, a modulus field $\phi$ and an axion $a$ coupled to $G\tilde{G}$ of the dark sector. The relevant Lagrangian is:
\begin{equation}\label{Dark_YM_lagrangian}
\mathcal{L}= \frac{a}{F_a}G\tilde{G} + y\phi\bar{\psi}\psi-V(\phi)\,,
\end{equation}
where the modulus potential is given by a fully general quartic potential:
\begin{equation}\label{eq:classical_potential}
V(\phi)=\kappa_1^3\phi-\frac{\mu}{2}^2\phi+\frac{\kappa_3}{3!}\phi^3+\frac{\lambda}{4!}\phi^4\,.
\end{equation}
The theta-term of the associated dark sector is given by:
\begin{equation}
\theta_{dark}=\theta_{bare}+\frac{a}{F_a}+arg[m_\psi]\,.
\end{equation}
The term $\theta_{bare}$ also includes possible contributions from light fermions in the dark sector. However, the existence of these additional dark light fermions is not relevant for the mechanism in which only the dark fermion $\psi$, coupled to the modulus $\phi$, is essential. 
This toy model is able to implement a shift of the sign of the fermion mass inducing the shift - by a factor $\Delta\theta=\pi$ -, of the axion potential between the inflationary epoch and today (see Fig.(\ref{fig: phi true})). 

When the dark sector confines, non-perturbative effects induce a potential for the axion field:
\begin{equation}\label{axion_dark_potential}
V(a)=\Lambda_D^4\left(1-\cos\left(\frac{a}{F_a}-\theta_{dark}\right)\right)\,.
\end{equation}
As explained in section \ref{sec_stochastic_axiverse}, we require a long period of low-scale inflation so that the axion relaxes to the minimum of the potential. After reheating, as we will see later, we require the temperature to be larger than the dark confinement scale, $T_{RH}>m_\psi\sim m_\phi\gg\Lambda_D$. In this case, the axion potential turns off and we can implement the shift of the sign of the dark fermion mass:
\begin{equation}
m_\psi\rightarrow-m_\psi\,.
\end{equation}
This in turn induces a shift to the potential by a factor $\Delta\theta=\pi$ when, at lower temperatures, the axion potential reappears. As a consequence, the axion field, that resided at the minimum during inflation, will be placed at the maximum of the potential before starting to oscillate. 

\subsection{Dynamics after reheating}
\begin{figure}[t!] \centering%
	\includegraphics[width=.75\textwidth]{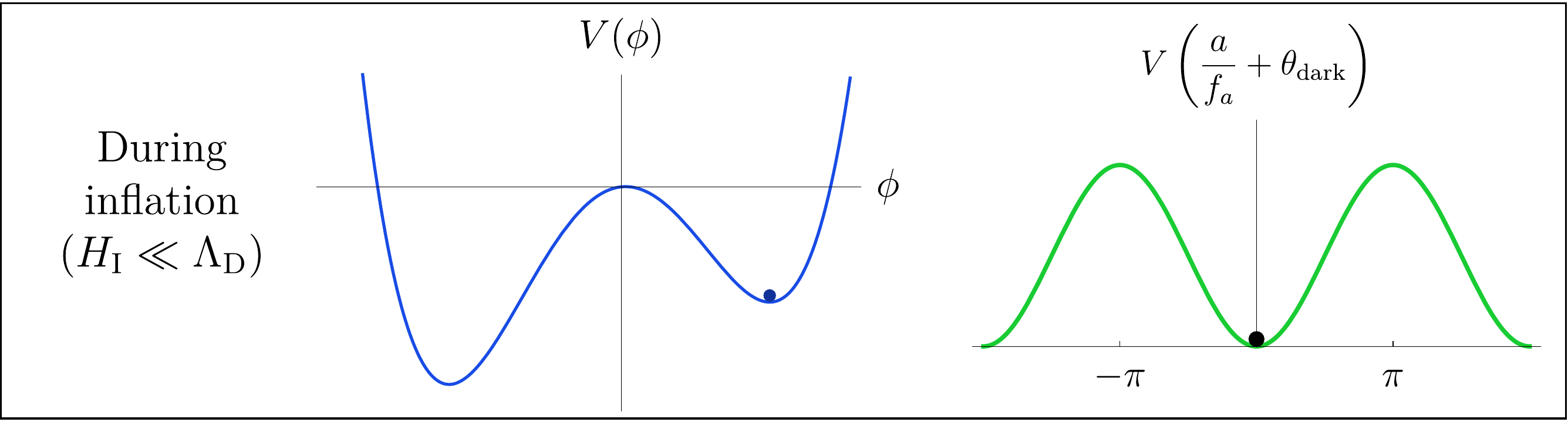}\smallskip\\%
	\includegraphics[width=.75\textwidth]{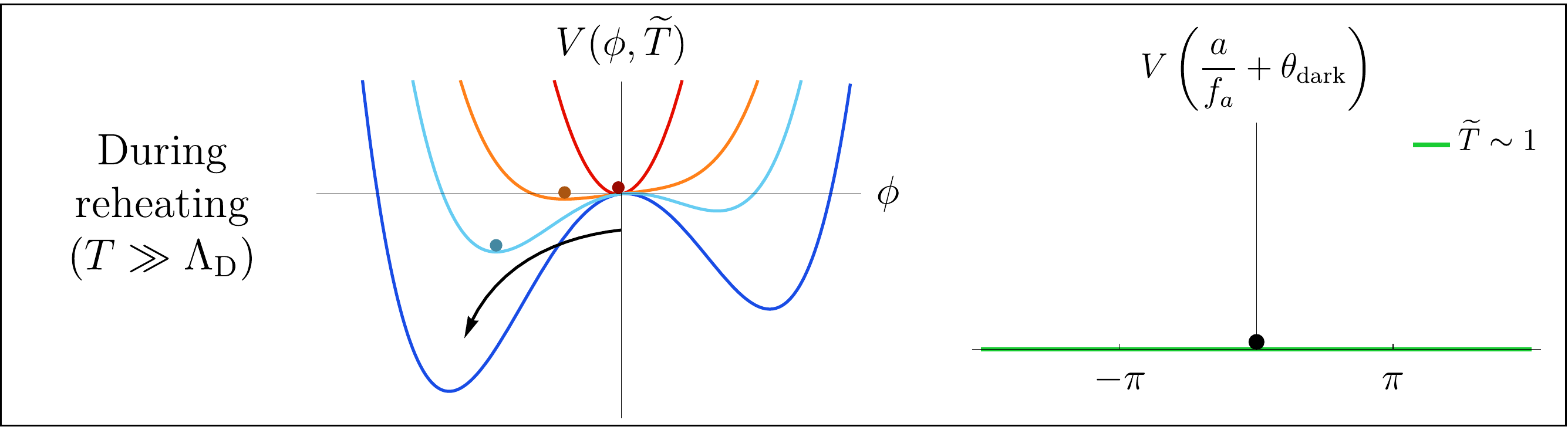} \smallskip\\%
	\includegraphics[width=.75\textwidth]{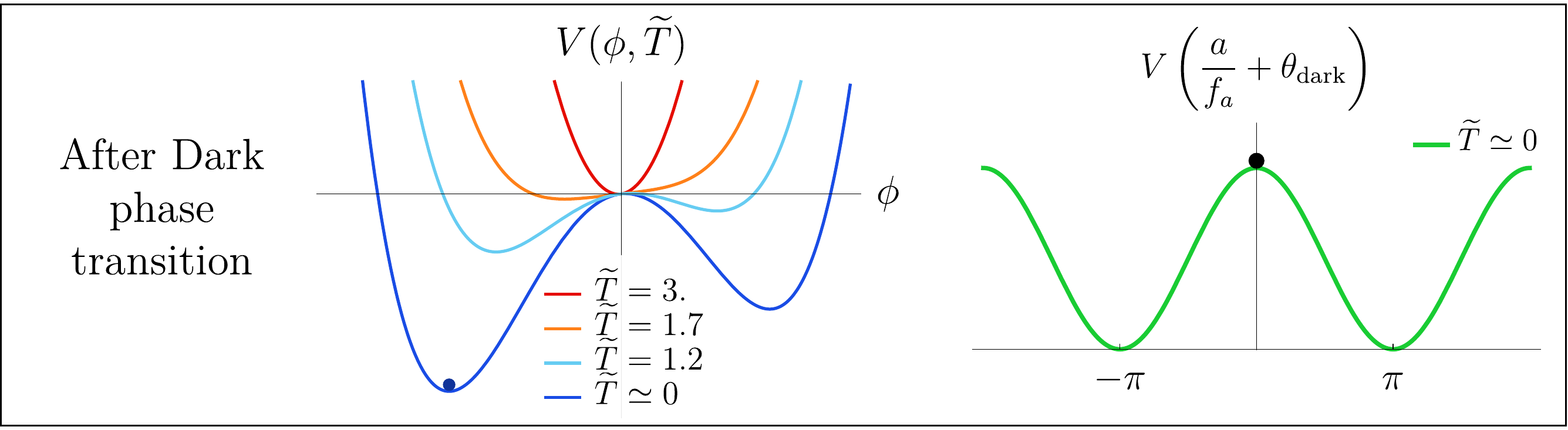} \smallskip\\%
	\caption{Dynamics for the modulus, $\phi$, and the axion field in the early Universe, adapted from \cite{Huang:2020etx}. Coloured lines show their potentials at different rescaled temperatures, defined as $\tilde{T}=T\, \mu^{-1}\lambda^{1/2}$, and the corresponding dots represent the field values at those times. In this case, the modulus lies in its false vacuum during inflation and rolls to its true vacuum after reheating. For simplicity, the Yukawa coupling is taken to be $y>0$. 
		\textbf{Top panel}: We assume that the inflationary Hubble rate is below the dark cofinement scale, $H_I<\Lambda_D$, so that the potential for the axion is turned on, and $a$ lies in the minimum (see Eq.~\eqref{axion_dark_potential}). 
		\textbf{Central panel}: During the thermal phase, the modulus field and the dark fermion, $\psi$, induce thermal corrections to the effective potential, $V(\phi)$, that lift the two vacua of $\phi$. 
		As the cosmic temperature decreases, the field $\phi$ rolls into the true vacuum.
		Due to the high cosmic temperature, during this period the axion potential is turned off. 
		\textbf{Bottom panel}: The modulus finally settles in its true vacuum. This minimum has opposite sign with respect to the inflationary vacuum. 
		The mass of the dark fermion, $\psi$, flips sign inducing a shift of $\pi$ in the argument of the dark fermion mass. 
		As a result, the axion potential is shifted by a factor of $\Delta\theta=\pi$, whereas the  initial value of the axion field has not changed since inflation. 
		The axion field sits at the top of its potential before the onset of oscillations and the initial misalignment is close to maximal up to small corrections due to the inflationary quantum fluctuations (see Eq.(\ref{small_misalignment})). 
	}%
	\label{fig: phi true}%
\end{figure}

In this section we summarize the vacuum selection mechanism that occurs after reheating, as the Universe cools down. We refer the reader to \cite{Huang:2020etx} for a detailed description of the vacuum selection process. Due to the interaction with the thermal plasma, for $T_{RH}> m_\phi$ the effective modulus potential has a single minimum around zero.  As the cosmic temperature decreases, we can always choose the minimum to which the modulus, $\phi$, rolls into in such a way that it is the opposite to the inflationary vacuum. This is done by studying the sign of the first derivative of the potential in Eq.(\ref{eq:classical_potential}) which is determined by the sign of the parameters $\kappa_1$ and $\kappa_3$.
This means that we can always arrange a combination of parameters, in Eq.(\ref{eq:classical_potential}), such that we flip the fermion mass,
\begin{equation}
m_\psi\rightarrow-m_\psi\,,
\end{equation}
and therefore induce a shift of the axion potential by a factor $\Delta\theta=\pi$, making the axion to start oscillating from the top of its potential. See Fig.\ref{fig: phi true} for a situation where the field $\phi$ settles in the true vacuum after reheating. Indeed, this vacuum selection mechanism does not require any tuning of the parameters of the modulus potential in Eq.(\ref{eq:classical_potential}) and, for an for an arbitrary inflationary vacuum -- that is, $\phi$ in a false or true inflationary vacuum state -- the probability that the axion starts from the top is around 50\% (see \cite{Huang:2020etx} for details of the available parameter space). The initial misalignment angle, taking into account inflationary quantum fluctuations, reads:
\begin{equation}
\theta_i=\pi-\delta\theta_i\,,\text{ with: }\,\delta\theta_i= \sqrt{\frac{3}{8\pi^2}} \frac{H_I^2}{\Lambda_D^2}\,.
\end{equation}
The main difference with \cite{Huang:2020etx} is that the scales: $m_\phi\sim m_\psi$, $\Lambda_D$, $T_{RH}$ will be rescaled to smaller values. This is because the typical potential heights that we will consider here, that is for ultralight axions and axion quintessence, are much smaller than the potential height of the QCD axion, $\Lambda_{QCD}^4$.  The core of the mechanism remains unchanged. 
\subsection{Non-YM contributions to the axion potential}
When the hidden sector is the only contribution to the axion potential, we have the following situation for the axion field: either it realises the maximal-misalignment mechanism and starts oscillating from the top, therefore resembling the situation studied in detail in \cite{Huang:2020etx}, or it does not realise the mechanism and its contribution to the energy density of the universe is negligible. As explained in section \ref{sec_stochastic_axiverse}, this occurs because the axion relic density is proportional to $\delta\theta^2$ and therefore extremely small when $H_I\ll O(100)$ eV. In other words, in our framework, instead of a continuous range for the initial misalignment angle we have a binary possibility: $\theta_i\simeq \pi$ or $\theta_i\simeq 0$ \cite{Huang:2020etx}.

A different possibility is that there is a sizeable source of the axion mass - that is, a source of shift symmetry breaking - which is not the YM we considered before but some string instanton or other non-perturbative effect that we cannot \textit{flip} with the mechanism above. This is, as we will see now, somewhat equivalent to the usual PQ quality problem of axion models. For simplicity, lets assume it to be of the form:
\begin{equation}
V=V_{\text{YM}}+V_{\text{new}}=\Lambda_D^4\left(1-\cos\left(a/F_a\right)\right)+\Lambda^{\prime\,4}(1-\cos\left(a/F_a-\delta\right))\,.
\end{equation}
The phase $\delta$ means that this new source of shift symmetry breaking, given by $V_{\text{new}}$, is not necessarily aligned with the hidden YM sector contribution that we can flip with the mechanism presented above. In this case, when we flip the YM but not the additional contribution to the potential, the axion does not start oscillating from the top because, during inflation, the minimum is at $\theta\simeq \frac{\Lambda^{\prime\,4}}{\Lambda_D^4}\sin\delta$. Of course this can only happen provided $H_I^4\ll \Lambda^{\prime\,4}$, with $\Lambda^\prime$ the scale of the non-YM shift symmetry breaking, so that this part of the potential is also turned on during inflation. In this case, the initial misalignment angle - that is, the offset with respect to the top - once we shift the hidden sector potential after inflation will be:
\begin{equation}\label{misalignment_non-YM}
\delta\theta\simeq k \frac{\Lambda^{\prime\,4}}{\Lambda_D^4}\sin\delta\,,
\end{equation}
with $k$ an $O(1)$ number and where we have assumed $\Lambda^\prime\lesssim\Lambda_D$. This gives an $O(1)$ initial misalignment angle if both scales, $\Lambda_D$ and $\Lambda^\prime$, are comparable.
Of course, when the new part of the axion potential is very small compared to the YM contribution, $\Lambda^\prime\ll\Lambda_D$, and/or $H_I\geq \Lambda^\prime$, then it can be neglected during inflation and the previous results apply, that is, the offset with respect to the top is small and given by $\delta\theta\sim H_I^2/\Lambda_D^2$.  
	
In the case that the non-YM contribution is comparable to the potential generated by the dark YM interaction but we do not flip any of them, the axion field will just roll to the minimum during inflation and will have a negligible contribution to the energy densiy of the Universe. Similarly, when the YM part is subdominant compared to the non-YM potential:
	\begin{equation}
	\Lambda^{\prime\,4}\gg \Lambda_D^4\,,	
	\end{equation}
the axion will just sit extremely close to the minimum during the inflationary period without any remarkable cosmological effect after reheating, as its abundance is very suppressed by $\delta\theta^2$, even if we flip the YM contribution. For simplicity, during the next sections we will just assume that this kind of non-YM contribution to the potential is sufficiently small and negligible in comparison to the dark YM that we can flip. We will consider this kind of effect later, in a string theory model building context, to show that they can be kept under control with reasonable assumptions.

\section{Applications: from ultralight axions to axion quintessence}
\label{sec:applications}
As explained above, our mechanism provides us the possibility of reducing the (in principle random) initial condition of the axion field to a binary choice:
\begin{equation}
\theta_i\simeq 0\,\,\text{ or }\,\,\theta_i\simeq\pi\,.
\end{equation}
Therefore, in the context of the axiverse, where several axions per decade of energy can appear, we can just \textit{switch on} the desired contribution of axions: e.g. those that compose the DM and/or DE we observe by coupling the corresponding axion to a YM sector as in Eq.(\ref{Dark_YM_lagrangian}) or an equivalent mechanism. The rest of dangerous axions and moduli, that is the ones that overclose the Universe or cause other related cosmological problems, are \textit{switched off} automatically after the long and low-scale inflationary period and do not harm the cosmic evolution after reheating. 

Additionally, as we will see now, there can be other interesting effects when we apply this mechanism - characterized by a low-scale and long inflationary period -, to the string axiverse. We have a mechanism which provides us with a dynamical explanation of a very particular initial condition of the axion field: it can start to oscillate from the top of its potential. When this occurs, as shown in \cite{Arvanitaki:2019rax}, oscillations are delayed with respect to the usual prediction of the misalignment mechanism. This means that $H_{osc}\ll m_a$, instead of the usual situation with $H_{osc}\sim m_a$. In \cite{Arvanitaki:2019rax}, the authors show that the axions which start oscillating from the top of their potential form some sort of dense DM clumps with diverse observational implications depending on their mass. This situation differs from the standard axion DM scenario. 

In addition to these implications studied in \cite{Arvanitaki:2019rax}, it is reasonable to ask if oscillations can be sufficiently delayed so that the axion field, in a slow-roll state, dominates the energy density of the Universe. In other words, if the axion field behaves as some sort of dark energy as a consequence of the maximal-misalignment mechanism. As we will now show, this seems indeed a reasonable question in the context of the axiverse where large decay constants, $M_{GUT}\lesssim F_a\lesssim M_P$, commonly appear. Assuming that the minimum of the axion potential has approximately zero energy (as in Eq.(\ref{axion_dark_potential})), the condition that the axion field, close to the top of its potential, dominates the energy density of the Universe reads:
\begin{equation}
2\Lambda_D^4\equiv 2m_a^2F_a^2\sim 3H^2M_P^2\,.
\end{equation}
This means that, in order to dominate the energy density of the Universe, the axion field has to lie close to the top of its potential, before starting to oscillate, for a time given by:
\begin{equation}
m_a\Delta t \simeq\sqrt{\frac{3}{8}}\frac{M_P}{F_a}\,,
\end{equation}
where we have assumed a radiation dominated Universe, that is $H=1/2t$, prior to dark energy domination. For large decay constants, $M_{GUT}\lesssim F_a \lesssim M_{P}$, which are expected in string theory constructions \cite{Arvanitaki:2009fg,Svrcek:2006yi}, we find that $\Delta t$ is larger than $m_a^{-1}$ by a factor $O(1-100)$.  Therefore, the effect occurs only if oscillations are sufficiently delayed. It is important to notice that this applies in general both for radiation and matter domination (with the appropriate prefactor for $H^{-1}$ when converted into time) prior to dark energy domination and also applies for an axion at any mass scale. 

When the axion field starts at the top of its potential we have to deal with a tachyonic instability. In other words, the concavity of the cosine potential close to the top generates a tachyonic instability of the axion field making it to grow exponentially with time \cite{Kaloper:2005aj,Dutta:2008qn}. The equation of motion for the small offset with respect to the top reads: 
\begin{equation}
\ddot{\delta a}+3H\dot{\delta a}-m_a^2\delta a=0\rightarrow \delta a(t)\simeq \delta a_i e^{mt}\,.
\end{equation}
The initial condition, $\delta a_i=F_a\delta\theta_i$, has therefore to compensate a potentially huge exponential growth of the field.

The axion field starts oscillating around the minimum when $\delta\theta (t)\sim O(1)$. If we want the axion to be at the top for a time $\Delta t$ prior to the onset of oscillations, then the initial offset with respect to the top cannot be larger than a maximum initial offset defined as:
\begin{equation}
\frac{\delta a_i}{F_a}\lesssim\delta\theta_{max}\sim e^{-m_a\Delta t}\,.
\end{equation}
This is the origin of the exponential tuning that in principle axion quintessence with $F_a\lesssim M_P$ requires to overcome the tachyonic instability. However, the maximal-misalignment mechanism we have introduced in the previous section provides us a typical offset: $\delta\theta_i\sim H_I^2/\Lambda_D^2$. Then, if the condition:
\begin{equation}\label{quint_condition}
\delta\theta_i\lesssim \delta\theta_{max}\rightarrow \sqrt{\frac{3}{8\pi^2}}\frac{H_I^2}{\Lambda_D^2} \lesssim e^{-m_a\Delta t}\,,
\end{equation}
is satisfied, the axion stays close to the top and behaves as vacuum energy for a time $\Delta t$, dominating the energy density of the Universe. After that time, it will start oscillating around the minimum with the consequent period of matter domination. Depending on the mass of the axion, this effect will be catastrophic or can be used to describe dark energy today. On the other hand, when Eq.(\ref{quint_condition}) is not satisfied, we have a situation where the axion starts to oscillate from close to the top of its potential with a behavior analogous to the situation studied in \cite{Arvanitaki:2019rax}. We explore these possibilities in the following sections.

As a summary, using the previous estimates, the axion dominates the energy density of the Universe in a slow-roll state if: 
\begin{equation}\label{radiation_condition}
\sqrt{\frac{3}{8\pi^2}}\frac{H_I^2}{\Lambda_D^2}\lesssim e^{-\sqrt{\frac{3}{8}}\frac{M_P}{F_a}}\,.
\end{equation}
Note that this effect is clearly unimportant for axions with decay constant $F_a\ll M_{GUT}$, but it can be certainly relevant for axions appearing in string constructions where:
\begin{equation}
10^{16}\text{ GeV}\lesssim F_a\lesssim 10^{18}\text{ GeV}\,,
\end{equation} 
are easily achieved. In the following we consider the possibility that this occurs for different axions. We consider, as benchmark scenarios, the cases of ultralight axion dark matter and axion quintessence with $m_a\sim O(1-100)H_0$ and decay constant well below the Planck scale. 

\subsection{Ultralight axions and \textit{heavy} axions}
\label{sec:ULA}
\begin{figure*}[t]
	\centering
	\includegraphics[width=0.6\textwidth]{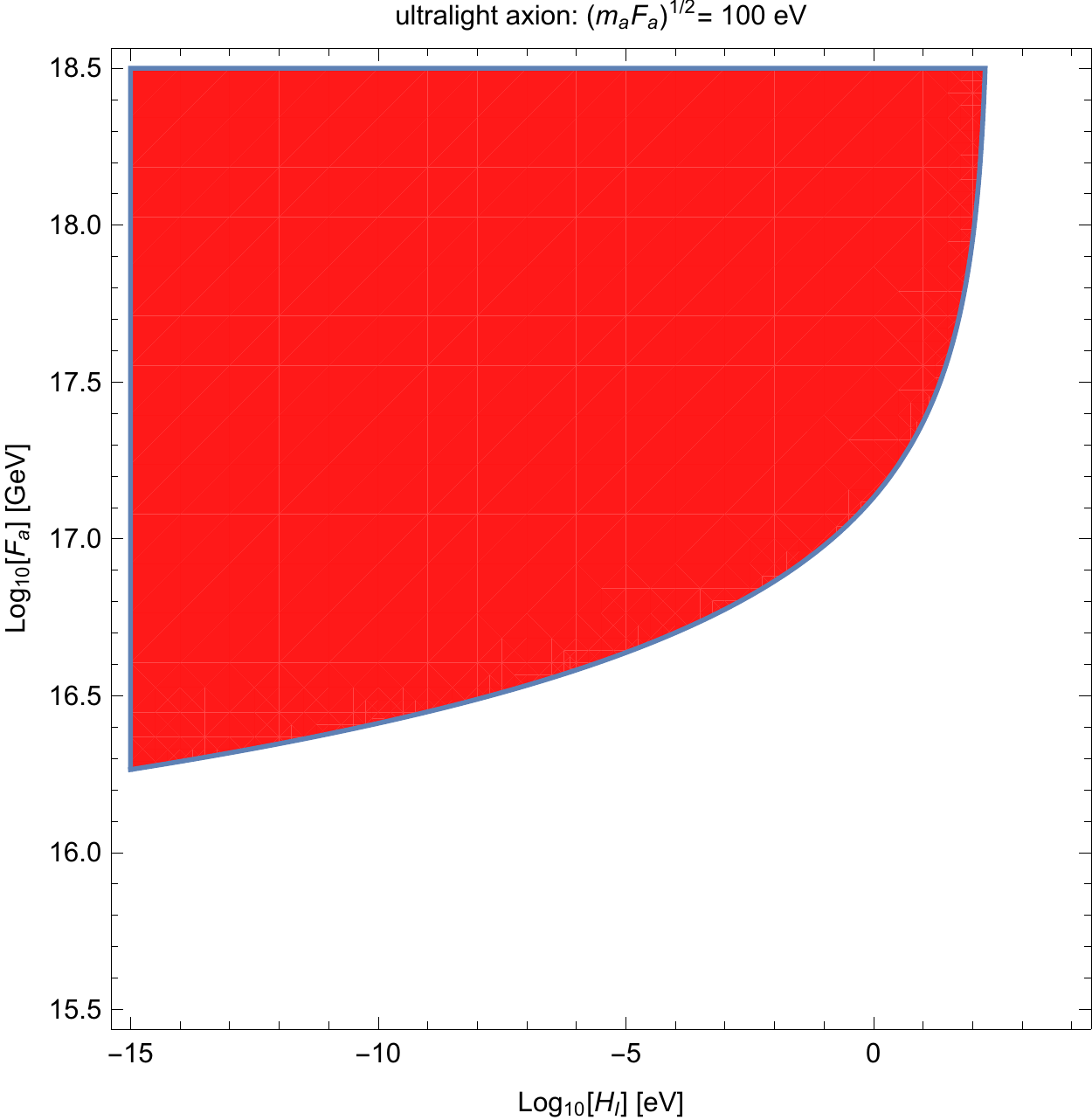}
	\caption{Parameter space allowed for maximal-misalignment in a ultralight axion model. In the red region, the ultralight axion behaves as dark energy prior to matter domination. Therefore, this region leads to a cosmology very different from ours and is excluded. Outside these regions, the ultralight axion has a misalignment close to maximal and can compose the observed abundance of DM. See text for details.}
	\label{fig_ULA}
\end{figure*}

Here we consider the case of ultralight axions and heavier axions such as the QCD axion. In the case of ultralight axions \cite{Hui:2016ltb}, we have an axion decay constant $F_a\sim 10^{17}$ GeV with a mass $m_a\sim 10^{-22}$ eV. This requires a dark confinement scale $\Lambda_D\sim 100$ eV if its potential has a YM origin. These axions easily appear in string compactifications \cite{Svrcek:2006yi}. In the standard misalignment case, the ultralight axion would start oscillating when $m_a\sim 3H $, this corresponds to a cosmic temperature of:
\begin{equation}
T_{osc}\sim (m_aM_P)^{1/2}\approx 500\,\text{eV}\,.
\end{equation}
Oscillations are delayed if the axion starts to oscillate from the top, reducing $T_{osc}$ by a factor $O(1-10)$. Since we are still well inside the radiation era, we can use Eq.(\ref{radiation_condition}) to estimate $\delta\theta_{max}$.  Therefore, in the case:
\begin{equation}
\delta\theta_i\lesssim \delta\theta_{max}\sim e^{-\sqrt{\frac{3}{8}}\frac{M_P}{F_a}}\,,
\end{equation}
the ultralight axion dominates the energy density of the Universe  behaving as vacuum energy. This corresponds to the red colored region in Fig.(\ref{fig_ULA}). Of course this would lead to a Universe very different from the one we observe, as it would evolve from a radiation dominated phase to dark energy domination which, after the ultralight axion oscillates, would change to a matter dominated phase. We believe, however, it illustrates the mechanism and might have diverse applications in cosmology. 

Of course, this is not the end of the story since depending on the decay constant $F_a$ the effect can be negligible. For example, a single ultralight axion with $m_a\sim 10^{-21}$ eV and $F_a\sim 10^{16}$ GeV could compose the totality of the DM if the initial misalignment is close to maximal, which is precisely the effect our mechanism is producing: $\delta\theta_i\sim H_I^2/\Lambda_D^2$. Note that despite such an ultralight axion never dominates the energy density of the Universe as vacuum energy before oscillating due to its decay constant (see Fig. (\ref{fig_ULA})), it can have the DM effects studied in \cite{Arvanitaki:2019rax}. This occurs because, as we have argued before, the initial offset with respect to the top is very small when $H_I\ll 100$ eV. This maximally-misaligned ultralight axion is not excluded and constitutes a well-motivated target for experimental searches.

We briefly comment on heavier axions such as the QCD axion or other axions appearing in the axiverse with $m_a\gg 10^{-22}$ eV. Note that the angle:
\begin{equation}
\delta\theta_{max}\sim e^{-\sqrt{\frac{3}{8}}\frac{M_P}{F_a}}\,,
\end{equation}
does not depend on the axion mass. Therefore, as long as $F_a\ll M_{GUT}$, any axion equipped with the maximal axion misalignment could be a good DM candidate without any worrisome contribution to early dark energy. Despite $\Lambda_D$ will be considerably larger than the ultralight axion case, the exponential suppression will easily make $\delta\theta_{max}$ smaller than the initial $\delta\theta_i\sim H_I^2/\Lambda_D^2$. The concrete value of $F_a$ allowed by the observed relic density of DM will be fixed once $m_a$ is specified. How easily these small decay constants, $F_a\ll M_{GUT}$, appear in string compactifications is a subject that we do not address here. Later, in section (\ref{sec:isocurvature}), we will see that if dark energy comes from a maximally-misaligned light axion, we get an upper bound to the DM axion mass from the isocurvature fluctuations bounds. 

\subsection{Dark energy and quintessence}
\label{sec:quintessence}
\begin{figure*}[t]
	\centering
	\includegraphics[width=0.6\textwidth]{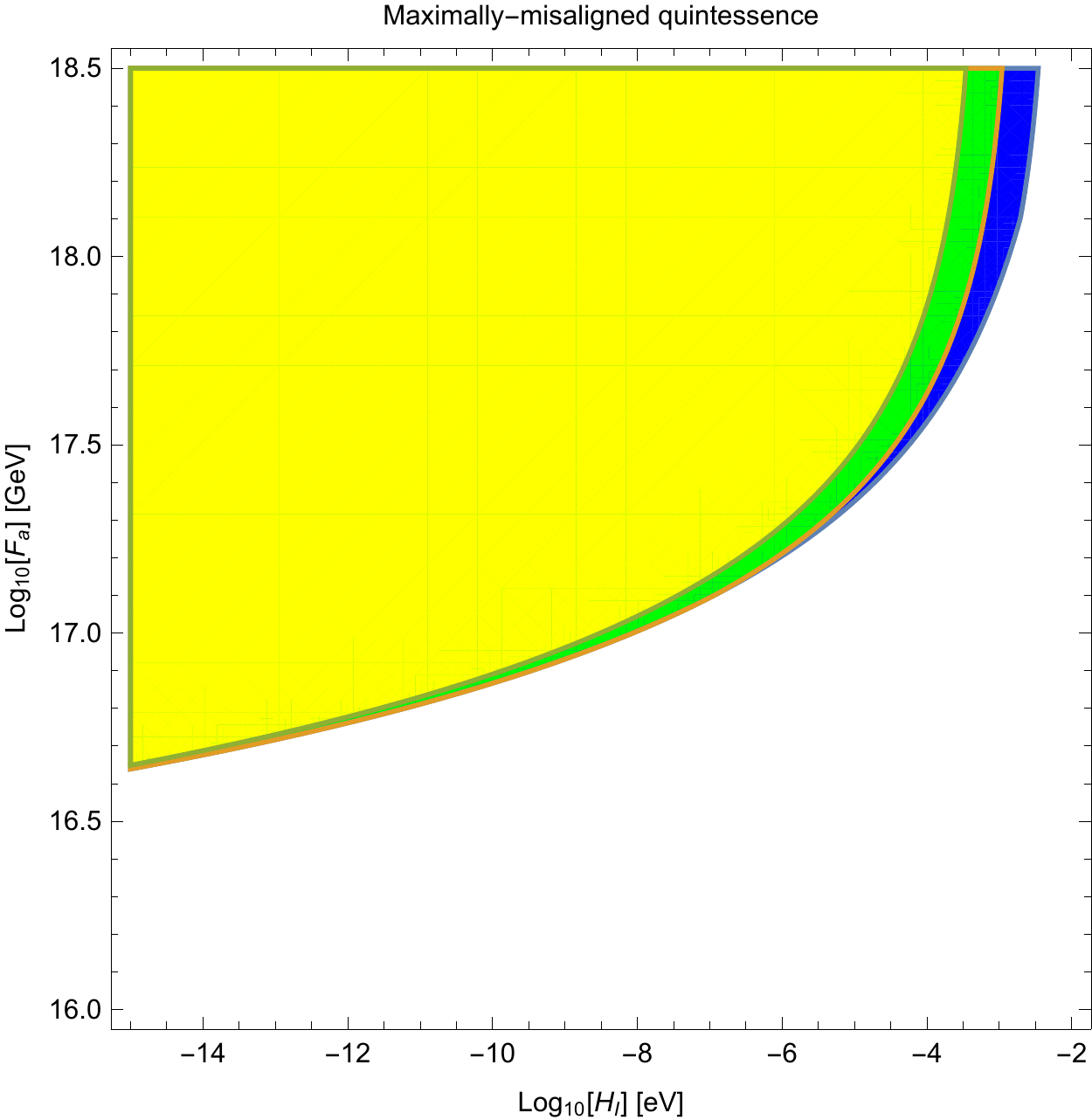}
	\caption{Parameter space allowed for (single axion) maximally-misaligned quintessence. The lower is $H_I$, the lower can be the axion decay constant $F_a$. The axion mass changes accordingly so that the relation $(m_aF_a)^{1/2}=2.5$ meV is maintained. The blue region satisfies $\Delta\theta(z=0)\lesssim 0.5$, the green satisfies $\Delta\theta(z=0)\lesssim 0.05$ and the yellow $\Delta\theta(z=0)\lesssim 0.005$. This traversed field distance fixes the dark energy equation of state today, $w(0)$ (see Eq.(\ref{EOS})). All the regions satisfy Planck 2018 bound, $w\lesssim -0.95$ \cite{Aghanim:2018eyx}.}
	\label{fig_quintessence}
\end{figure*}

We now consider the case of axions with a mass comparable to the Hubble parameter today $H_0$. It is well-known that these axions can describe dark energy, i.e., an  accelerated expansion in our Universe after a period of matter domination. In this case, assuming that a single light axion is the responsible of dark energy today,
\begin{equation}
2m_a^2F_a^2\sim \left(\Omega_\Lambda\right) 3H_0^2M_P^2\,\rightarrow\,m_aH_0^{-1}\simeq M_P/F_a\,.
\end{equation}
%
%As previously anticipated, we can use the maximal-misalignment mechanism to have natural theories of quintessence where the axion field has an initial condition $\theta_i\approx\pi$. The potential height has to be equal to the energy density of dark energy:
%%
%\begin{equation}
%2\Lambda_D^4\simeq\left(\Omega_\Lambda\right)3 H_0^2M_P^2\,,\rightarrow\,m_aH_0^{-1}\simeq M_P/F_a\,.
%\end{equation}
%
Therefore, the previous estimate (see Eq.\ref{radiation_condition}) changes to:
\begin{equation}\label{quintessence_condition}
e^{-M_P/F_a}\geq \sqrt{\frac{3}{8\pi^2}}\frac{H_I^2}{\Lambda_D^2}\,,
\end{equation}
where we impose $\Lambda_D\simeq 2.5$ meV to have the appropriate value of dark energy density:
\begin{equation}
2\Lambda_D^4\simeq\rho_{vac}=(3\,\text{meV})^4\,.
\end{equation}
The above equation (\ref{quintessence_condition}) is telling us that one can have a consistent quintessence model from a single axion with sub-Planckian decay constant, $F_a\ll M_P$ and $m_a\gg H_0$, provided $\Lambda_D\gg H_I$. The region of parameter space satisfying this relation is in Fig.\ref{fig_quintessence}.

We can go a step further and estimate the equation of state as a function of the redshift:
\begin{equation}
1+w(z)\simeq \frac{(F_a\dot{\theta})^2}{V}\,.
\end{equation}
Assuming that the field is slowly rolling at present we have: 
\begin{equation}\label{EOS}
1+w(z)\simeq \frac{V^\prime(a)^2}{9H^2V}\approx\frac{1}{18}\frac{m_a^2}{H_0^2}\Delta\theta^2\,,
\end{equation}
with $\Delta\theta$ the distance that the axion field has rolled. Notice that we need to impose $\Delta\theta<1$ before it starts oscillating. The smaller is the distance the axion has rolled, the closer we are to the cosmological constant prediction,  $w_\Lambda=-1$. For example, the yellow region in Fig.(\ref{fig_quintessence}) satisfies $\Delta\theta\leq0.005$ today and, therefore, gives typical values of $1+w(0)\ll 10^{-3}$ for decay constants slightly below $F_a\sim 10^{17}$ GeV.

\subsection{The maximally-misaligned axiverse and $N$-essence}
\begin{figure}[t]
	\centering
	\includegraphics[width=0.6\textwidth]{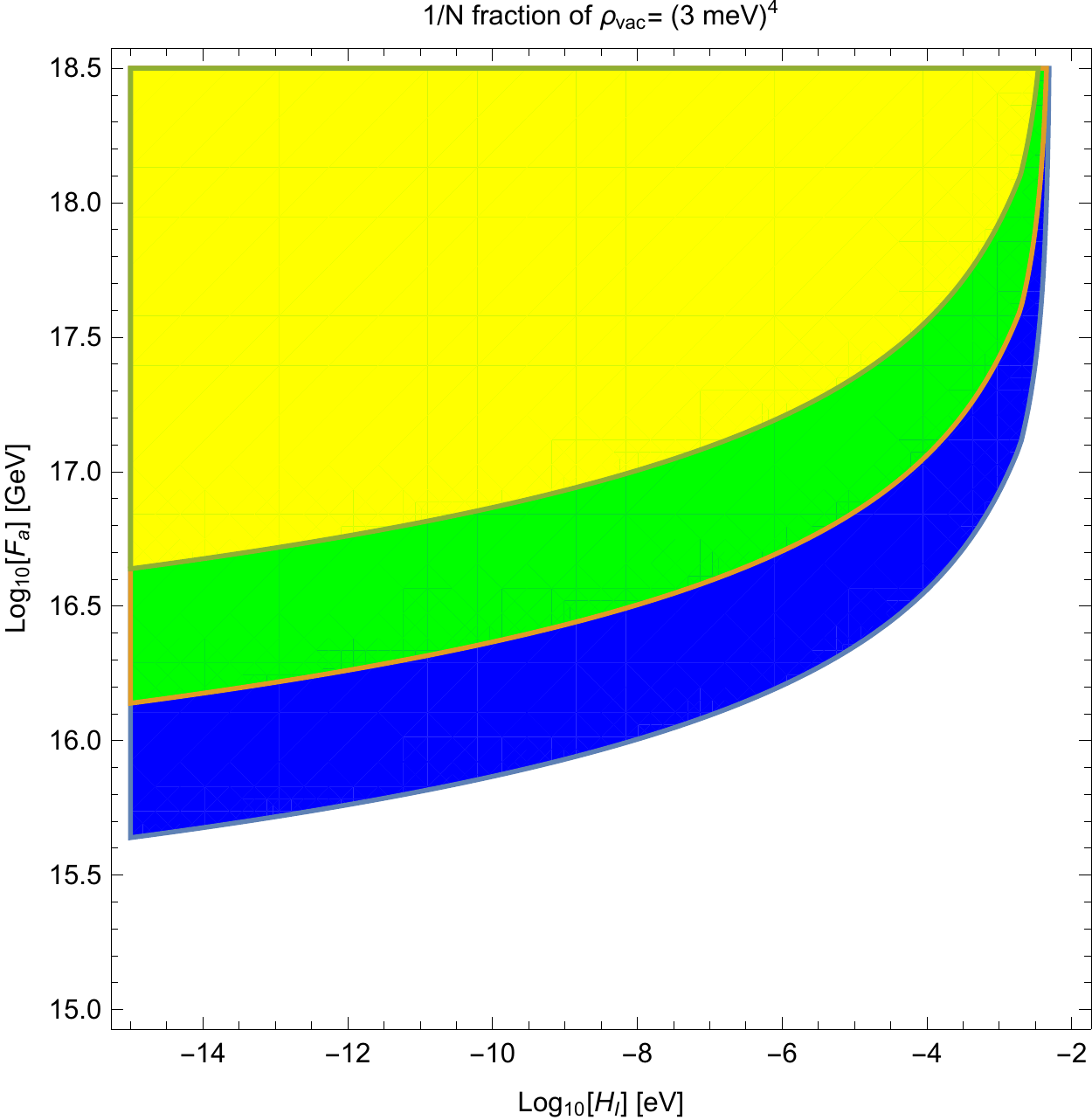}
	\caption{Parameter space allowed for $N$-essence, defined as $\delta\theta_i\lesssim \delta\theta_{max}$ (see Eq.(\ref{condition_N-essence})). Different colors correspond to light axions composing a different fraction of the observed dark energy. The lower is $H_I$, the lower can be the axion decay constant $F_a$. The yellow region corresponds to $N=1$, the green $N\leq10$, the blue $N\leq 100$. We see that a moderate number of axions $N=O(10)$ already allows a decay constant $F_a\sim M_{GUT}$ and an inflationary Hubble parameter $H_I\sim O(0.1-1)$ meV. Assuming instantaneous reheating, this corresponds to a reheating temperature of $T_{RH}\sim O(0.1-1)$ TeV, i.e. around the weak scale. As in the case of single axion quintessence, the plotted regions satisfy the current bound to the equation of state parameter: $w\lesssim -0.95$ \cite{Aghanim:2018eyx}.
}
	\label{fig_Nessence}
\end{figure}

In addition to axion quintessence dominance of the dark energy density, there is always the possibility that maximally-misaligned quintessence constitutes only a subdominant part of dark energy on top of some other type of vacuum energy (e.g. cosmological constant): 
\begin{equation}
2m_a^2F_a^2\sim \left(\frac{\Omega_\Lambda}{N}\right)3H_0^2M_P^2\,.
\end{equation}
If $N=100$, then quintessence is only $O(1)$\% of dark energy and still could have observable effects \cite{Minami:2020odp,Fujita:2020ecn}. The condition now reads:
\begin{equation}\label{condition_N-essence}
e^{-M_P/(\sqrt{N}F_a)}\geq \sqrt{\frac{3}{8\pi^2}}\frac{H_I^2}{\Lambda_D^2}\,.
\end{equation}
For a fixed axion mass, the decay constant we need to satisfy the condition is reduced by a factor $\sqrt{N}$ with respect to the standard estimate of Eq.(\ref{quintessence_condition}). See different colors in Fig.(\ref{fig_Nessence}).

A third possibility is that several axions realise the maximal-misalignment mechanism and their collective contribution results in the observed dark energy density. Here we extend our mechanism for the case of $N$ axions showing that it shares some of the benefits of qui$\mathcal{N}$tessence scenarios \cite{Kaloper:2005aj} with a moderate number of axions, $N$. By analogy we call this scenario $N$-essence. 

We consider, as in \cite{Kaloper:2005aj}, a collection of $M$ non-interacting\footnote{Turning on interactions with other axions (cross terms at the axion potential) generically leads to new directions in field space that, in some cases, might result in a faster rolling, spoiling the success of maximally-misaligned quintessence. Given that the details of the cross terms are rather model dependent, and in a string theory context strongly depend on details of the compactification manifold (see \cite{Mehta:2021pwf}), we will consider only the case of non-interacting axions.} light axions:
\begin{equation}
V=\sum_i^M \Lambda_n^4(1-\cos(a_i/F_{a_i}))\,.
\end{equation}
In the case there are $N\leq M$ light axions realising the maximal-misalignment mechanism, contributing as vacuum energy, the total energy density is just the sum of the individual contributions, which we impose to match the observed dark energy density:
\begin{equation}
\rho_{vac}=\sum_n^N\rho_n=\left(\Omega_\Lambda\right)3 H_0^2M_P^2\,.
\end{equation}
Assuming, for simplicity, that the height of their potential is comparable (for simplicity,  $m_{a,i}\sim m_{a}$ and $F_{a,i}\sim F_{a}$) we get as expected:
\begin{equation}
m_a/H_0\simeq\frac{1}{\sqrt{N}}\frac{M_P}{F_a}\,.
\end{equation}
As before, this means that, for $m_a$ fixed, we can further reduce the required decay constant $F_a$ by a factor $\sqrt{N}$. The difference is that now the collective effect of $N$ axions describes the totality of the observed dark energy density. As shown in Fig. (\ref{fig_Nessence}), in the case of the $N$-essence scenario, we can reduce the required decay constant down to $F_a\sim 10^{16}$ GeV just by taking $N\sim O(10)$ maximally-misaligned axions. Without any dynamical mechanism of maximal-misalignment -- that is, a flat distribution for the initial misalignment, $\theta_i$ -- the chances of having $N\sim O(10)$ axions with $\theta_i\sim\pi$ would be vanishingly small. However, in the light of the mechanism presented in Sec. \ref{sec:toy_model} every axion, $a_i$, has a $\sim 50\%$ probability \cite{Huang:2020etx} to start from the top of its potential regardless of the inflationary vacuum of the associated modulus, $\phi_i$.

Another comment is in order. As we remarked above, it is most likely that only a few axions of the axiverse implement the maximal-misalignment mechanism. In this case, for example, we could have the handful of axions corresponding to $N$-essence and one or more axions corresponding to ultralight axion DM with mass(es) around $m_a\sim 10^{-22}$ eV or larger. One could ask about the fate of the others. In order to explain dark energy with a collection of maximally misaligned light axions, the mechanism requires a low Hubble parameter during inflation:
\begin{equation}
H_I\lesssim \rho_{vac}^{1/4}\sim 3\text{ meV}\,.
\end{equation}
As already anticipated in section \ref{sec_stochastic_axiverse}, this automatically implies that all the \textit{phenomenologically relevant} axions must have relaxed to their minimum during inflation. Therefore only those featuring the maximal-misalignment, or a similar mechanism, can have observable effects\footnote{Superradiance and the rotation of CMB polarization (for axions with $m_a\lesssim H_0$) are exceptions to this.}. Axions that lie at the minimum of the potential,  with a contribution suppressed by $ \delta\theta^2\sim H_I^4/(m_aF_a)^2$,
do not contribute appreciably to the energy density of the Universe. In fact, for a given $H_I$, the heavier - and in principle the more dangerous - is the axion, the smaller is the spread of the distribution around the minimum and the smaller is its contribution to the energy density budget of the Universe (see Eq.(\ref{relic_small_misalignment})). On the other hand, axions with $(m_aF_a)\ll H_I^2$ will not relax to the minimum as their potential was not turned on during inflation. However, this also implies that their contribution is:
\begin{equation}
\rho_{light}\sim (m_aF_a)^2\ll \rho_{vac}\,,
\end{equation}
so they appear to be harmless from the point of view of cosmology. In fact, for reasonable decay constants they can be safely considered massless. The maximal-misalignment mechanism, in combination with a low inflationary Hubble scale, suggests itself as a compelling combination to avoid axion overclosing of the Universe and other related issues of the axiverse cosmology while having a consistent description of dark matter, dark energy and several interesting signals such as BH superradiance or the rotation of CMB polarization. 

\section{Phenomenological implications}
\label{sec:pheno}
We require that axions that may overclose the Universe are \textit{regulated} by the long inflationary period and roll to their minimum. The scenario is therefore characterized by a low inflationary Hubble scale, typically $H_I\lesssim 100$ eV. In this case, axions with
\begin{equation}
 (m_aF_a)^{1/2}\ll H_I\lesssim 100 \text{ eV}\,,
\end{equation}
will of course have random initial conditions. However they generically do not pose a serious problem for cosmology. 

In order to describe dark energy with maximally misaligned quintessence we require:
\begin{equation}
H_I\lesssim \Lambda_D\sim \rho_{vac}^{1/4}\simeq 3 \text{ meV}\,.
\end{equation}
There is a lower bound to the inflationary Hubble parameter which comes from the fact that the reheating temperature has to be sufficiently high to allow successful BBN at temperatures of the order of the MeV scale \cite{deSalas:2015glj}. Assuming instantaneous reheating, the maximal temperature reads:
\begin{equation}
T_{RH}\sim\left(\frac{90}{g_\star\pi^2}\right)^{1/4}\sqrt{H_IM_P}\sim 30 \text{ TeV}\left(\frac{H_I}{\text{eV}}\right)^{1/2}\gg\vev{\phi}\,.
\end{equation}
The last inequality is required by the consistency of the mechanism in section \ref{sec:toy_model}. Requiring $T_{RH}>O(1)\,\text{MeV}$, implies:
\begin{equation}
H_I\gg 10^{-14}\text{ eV}\,.
\end{equation}
This low reheating temperature has potentially observable effects as we will see now. 
\subsection{$\Delta N_{eff}$}
We have considered axions with a potential coming from a dark YM interaction. The associated confinement scales are relatively small: $\Lambda_D\sim 100$ eV for the case of an ultralight axion and $\Lambda_D\sim 3$ meV for axion quintessence. In either case, the associated dark-gluons will be relativistic during most of the evolution of the Universe and will act as radiation during BBN, contributing to $\Delta N_{eff}$. In addition, the new Dirac fermion and modulus might also contribute depending on their mass. Despite in most of the parameter space their mass is larger than $T_{BBN}$, we will take a conservative approach and consider that, in addition to the dark gluons, they also contribute to the effective number of neutrinos. 

The energy density of the new non-abelian dark sector is related to the effective number of neutrinos through \cite{Buen-Abad:2015ova}:
\begin{equation}
\Delta N_{eff}=\frac{\rho_D}{\rho_\nu}=\frac{8}{7}g_\star^{D}\left(\frac{T_D^4}{T_\nu^4}\right)\,,
\end{equation}
where $g_\star^{D}$ is the number of relativistic d.o.f. of the dark sector and $T_D$, $T_\nu$ are the temperatures of the dark sector and neutrinos, respectively. If we consider that the dark gauge group is $G_D=SU(N)$, then we have:
\begin{equation}\label{Delta-N_eff}
\Delta N_{eff}=\frac{8}{7}\left[(N^2-1)+1+\frac{7}{8}\times 2\right]\left(\frac{T_D^4}{T_\nu^4}\right)\leq 0.33\,,
\end{equation}
where $\Delta N_{eff}\leq 0.33$ corresponds to the current limit from primordial $^4$He abundance at 95\% C.L \cite{Pitrou:2018cgg,Fields:2019pfx,Allahverdi:2020bys}. Even for the smallest non-abelian group, $G_D=SU(2)$, this clearly indicates the need of a dark sector colder than the visible sector:
\begin{equation}
\Delta N_{eff}=6.57 \left(\frac{T_D^4}{T_\nu^4}\right)\leq 0.3\rightarrow \frac{T_D}{T_\gamma}=(4/11)^{1/3}\left(\frac{T_D}{T_\nu}\right)\leq 0.32\,.
\end{equation}
This could be explained, for example, if there is a preferential reheating for our visible sector through a more efficient coupling to the inflaton sector. 

The previous estimate corresponds to the ratio of temperatures at BBN. We now estimate the ratio of temperatures of both sectors at the time of reheating. Assuming entropy conservation in each sector \cite{Feng:2008mu},
\begin{equation}
\frac{g_{\star, s}^D(T^D_{BBN}) T^{D\,3}_{BBN}}{g_{\star, s}^D(T^D_{RH}) T^{D\,3}_{RH}}=\frac{g_{\star, s}^{vis}(T_{BBN}) T^{3}_{BBN}}{g_{\star, s}^{vis}(T_{RH}) T^{3}_{RH}}\,,
\end{equation}
we relate the temperatures of the visible and hidden sector (denoted with a \textbf{"}D") right after reheating: 
\begin{equation}
\frac{T^D_{RH}}{T_{RH}}=\left(\frac{g_{\star, s}^{vis}(T_{RH})}{10.75}\right)^{1/3}\left(\frac{g_{\star, s}^{D}(T_{BBN}^D)}{g_{\star, s}^D(T^D_{RH})}\right)^{1/3}\frac{T^D_{BBN}}{T_{BBN}}\,.
\end{equation}
Assuming a relatively simple dark sector with few degrees of freedom, $g_{\star, s}^D(T^D_{RH})\sim g_{\star, s}^D(T^D_{BBN})$,  and $g_{\star, s}^{vis}(T_{RH})=106.75$, the previous limit translates into:
\begin{equation}
\frac{T^D_{RH}}{T_{RH}}\lesssim 0.55\,.
\end{equation}
The expectation is that CMB-S4 \cite{Abazajian:2016yjj} will improve the limit on the effective number of neutrinos to:
\begin{equation}
\Delta N_{eff}\leq 0.06\,\,\,\text{at $95\%$ C.L.}
\end{equation}
If the dark-gluons are still relativistic at around $T_{CMB}$, we can repeat the previous estimation and obtain the ratio:
\begin{equation}
\frac{T^D_{RH}}{T^{vis}_{RH}}\lesssim 0.35\,.
\end{equation}
One can easily see from Eq. (\ref{Delta-N_eff}) that,  for larger $SU(N)$ groups, the ratio of temperatures goes as $T^D/T^{vis}\propto 1/\sqrt{N}$. In the case there are $M$ $SU(N)$ hidden sectors coupled to ultralight axions then we have to rescale the previous result. The effective number of neutrinos becomes: 
\begin{equation}\label{Delta-N_eff_Msectors}
\Delta N_{eff}=M\times\frac{8}{7}\left[(N^2-1)+1+\frac{7}{8}\times 2\right]\left(\frac{T_D^4}{T_\nu^4}\right)\leq 0.3\,.
\end{equation}
Therefore, for $M$ SU(2) dark sectors, the bounds scales as:
\begin{equation}
\frac{T^D_{RH}}{T^{vis}_{RH}}\lesssim \frac{0.55}{M^{1/4}}\,.
\end{equation}
We see that unless the number of hidden sectors $M$ is large, the ratio of temperatures at the reheating scale is around $O(0.1)$. 
\subsection{Glueballs and other light, stable states}
Due to the presence of non-abelian confining interactions, the lightest unconfined states are stable. In the case there are no light fermions, the situation is similar to a pure YM scenario where the lightest states are glueballs \cite{Boddy:2014yra,Soni:2016gzf}. See also \cite{Halverson:2016nfq,Halverson:2018olu} for recent studies in a string theory context. In principle these glueballs tend to overclose the universe and their relic density can be estimated as:
\begin{equation}
\Omega_{gb} h^2\simeq \frac{\Lambda_D}{3.6\text{ eV}}\xi^3\,,
\end{equation}
where $\xi\sim (g_{\star, s}^D/g_{\star, s}^{vis})^{1/3}T_D/T_{vis}\,$ is the ratio of temperatures of the dark and visible sectors. 

Typical dark sectors that we employed in the maximal-misalignment mechanism have a confinement scale ranging from $\Lambda_D\lesssim 100$ eV for ultralight axion DM, to $\Lambda_D\sim\rho_{vac}^{1/4}\sim 3$ meV for the case of maximally-misaligned quintessence. The latter is not problematic from the point of view of glueball abundance but the former could be worrisome depending on $\xi$. The glueball abundance is further reduced by the need of a colder dark sector to satisfy the constraints on $\Delta N_{eff}$. For $\xi\lesssim 0.1$, it is expected that the energy density in glueballs is negligible even for dark sectors with $\Lambda_D\sim 100$ eV. In addition, the dark sector glueballs can mix with the associated axions. However, being extremely light this mixing is expected to be small. 

\subsection{Isocurvature fluctuations}
\label{sec:isocurvature}
The size of the quantum fluctuations of the axion field that we can observe in the CMB is given by:
\begin{equation}
\delta a\sim \frac{\sqrt{N_{CMB}}}{2\pi}H_I\,,
\end{equation}
with $N_{CMB}\simeq 8$. In principle these fluctuations are tiny for low-scale inflation, as it is in our case. However, they can get enhanced if the initial misalignment angle is close to $\theta_i\simeq \pi$ \cite{Kobayashi:2013nva}. Large anharmonic effects boost the impact of these perturbations which can be estimated as \cite{Arvanitaki:2019rax,Huang:2020etx}:
\begin{equation}
\delta_{\text{iso}}=\frac{\delta\rho}{\rho}\simeq C\frac{\Delta_{\delta\theta}}{\delta\theta|\ln\delta\theta|}\,,
\end{equation}
where $C\approx 1.5-2.5$ is approximately constant with $\delta\theta$. In our mechanism, the typical offset does also depend on $H_I$:
\begin{equation}
\delta\theta\sim \sqrt{\frac{3}{8\pi^2}}\frac{H_I^2}{m_aF_a}\,,
\end{equation}
This fact generates an interesting non-trivial dependence of $\delta_{\text{iso}}$ on $H_I$. As stated above, the size of typical fluctuations of the misalignment angle is given by:
\begin{equation}
\Delta_{\delta\theta}\simeq \frac{\sqrt{N_{CMB}}}{2\pi}\frac{H_I}{F_a}\,,
\end{equation}
and, therefore, we get:
\begin{equation}
\delta_{\text{iso}}\simeq\frac{C\sqrt{N_{CMB}}}{2\pi|\ln\delta\theta|}\sqrt{\frac{8\pi^2}{3}}\frac{m_a}{H_I}\,,
\end{equation}
where we leave $\ln\delta\theta$ explicit as it will take values, at most, of $O(10)$. The Planck experiment data requires the isocurvature perturbations to be subdominant by a factor $\sim 10^{-2}$ with respect to the adiabatic fluctuations \cite{Akrami:2018odb}. With this in mind we can obtain an upper bound for the DM axion mass with respect to the inflationary Hubble scale:
\begin{equation}
\delta_{\text{iso}}\lesssim\delta^{iso}_{\text{Planck}}\sim 10^{-5.5}\,\,\rightarrow\,\, m_a\lesssim 10^{-5}H_I\,.
\end{equation}
It is interesting to see that for the typical inflationary Hubble scales consistent with maximally-misaligned quintessence,  $H_I\lesssim 10^{-3}$ eV, the axion mass is required to be below $m_a<10^{-8}$ eV. This suggests that the QCD axion may not be the (dominant) DM candidate in the scenario of maximally-misaligned quintessence unless its decay constant is $F^{QCD}_a>10^{15}$ GeV. However, it leaves unconstrained typical light axion DM candidates such as the ultralight axions of the fuzzy DM scenario with axion masses around $m_a\sim 10^{-22}$ eV and $F_a\sim 10^{16-17}$ GeV. 

\section{Connections to String Theory model building}
\label{sec:string_model_building}
As we have seen in Sec.\ref{sec:toy_model}, maximally-misaligned axions are very sensitive to non-YM contributions to the axion potential. In addition to YM sectors, realistic string constructions are expected to come with additional sources of shift symmetry breaking. In this section we consider these effects and study the situation in which these non-YM instantons are kept under control. For the sake of completeness we consider a D-brane model in type-II string theories. Non-abelian gauge couplings arising from gauge sectors on D(3+q)-branes are inversely proportional to the volume of the q-cycle, $Q$, that a given D-brane wraps:
\begin{equation}\label{string_gauge_coupling}
\frac{1}{\alpha}=\frac{V_Q}{g_s l_s^q}\equiv\tau\,.
\end{equation}
In these scenarios the four dimensional Planck mass is given by:
\begin{equation}
M_P^2=4\pi\frac{V_X}{g_s^2l_s^8}\,,
\end{equation}
where $X$ is a compact manifold which, in addition to $Q$, also includes purely \textit{gravitational dimensions}. Depending on their number, if the purely gravitational dimensions are large compared to the fundamental string length $l_s$ (for example, $R^{-1}\sim$ MeV), one can easily have a low string scale not far from the TeV scale.
This is in fact desirable in our framework of maximally-misaligned axions, given that the highest reheating temperatures we can achieve are generically low. To avoid overclosure we need $H_I\lesssim 100$ eV which, translated into reheating temperatures (and assuming instantaneous reheating) gives $T_{RH}\lesssim 300$ TeV. In this case, in the absence of an alternative mechanism to reduce the tuning of the modulus mass $m_\phi$ (see Eq.(\ref{eq:classical_potential})), one can arrange the size of the extra dimensions in such a way that $l_s^{-1}\sim 10^{3-4}$ TeV\footnote{The issue of finding a consistent moduli stabilization mechanism is a separate issue which is interesting by itself and we do not address here.}.

In the most radical case one needs $H_I\lesssim O(1)$ meV to have maximally-misaligned quintessence. This gives reheating temperatures of order $T_{RH}\sim O(1-100)$ GeV (see Fig.(\ref{fig_quintessence})) for the case of a single quintessence field describing dark energy today. If there are $N\sim O(10)$ axions contributing to vacuum energy then we can have a substantially larger inflationary Hubble parameter, leading to $T_{RH}\sim O(0.1 - 1)$ TeV, and still be compatible with $F_a\sim M_{GUT}$ (see Fig. \ref{fig_Nessence}). In either case, when the reheating temperature is related to the size of the extra dimension as:
\begin{equation}
T_{RH}\sim m_\phi\sim m_\psi\lesssim R_{dark}^{-1}\,,
\end{equation} 
the tuning of the modulus mass, $m_\phi$, is minimized. Of course there may exist alternative mechanisms to reduce the tuning of the modulus mass, here we just describe the situation for illustrative purposes. In the absence of such a mechanism, the minimization of fine-tuning in our framework of maximally-misaligned quintessence suggests a low string scale, $M_s\sim l_s^{-1}$, not far from the $O(10)$ TeV scale.

Regarding the decay constant, as explained in \cite{Svrcek:2006yi}, in this kind of string theories a typical axion has a decay constant:
\begin{equation}
F_a\sim \frac{\alpha_D}{2\pi} M_P\,,
\end{equation}
which correspond to a decay constant around the GUT scale. Of course there exists the possibility of substantially smaller decay constants, in situations where the overall volume of the manifold is exponentially large in string units (see \cite{Broeckel:2021dpz} for a recent study) or by employing highly warped extra dimensions. We do not consider these cases further.

It is not implausible that such a light axion couples to photons. This is required to produce some of the signals mentioned above such as the rotation of CMB polarization. In order to remain light, we need to avoid the coupling to QCD or any other confining interaction that could give a mass larger than the hidden sector mass that we can flip. The requirement of remaining light also bounds the coupling constant of $U(1)_Y$ or $SU(2)_L$  at the compactification scale. The reason is that the gauge bosons of at least one of them is necessarily coupled to the maximally-misaligned axion if we want a sizeable coupling to photons at low energies. We consider this possibility in the next section.
\subsection{Protection from stringy effects}
%
%There are plenty of instantons and non-perturbative effects that contribute to the axion potential in string theory. 
Maximally-misaligned axions, and in particular for the axion quintessence case, are extremely sensitive to these additional sources of shift symmetry breaking due to the strong dependence on the initial conditions. Consistency of the mechanism presented in Sec. \ref{sec:quintessence} requires that the YM contribution - that is, the part of the axion potential that can be \textit{flipped} throughout the cosmic envolution -, dominates so that $\theta_i=\pi-\delta\theta$ is not \textit{polluted} by the effects of additional contributions to the axion potential. Typically, stringy instantons contribute to the axion potential as: 
\begin{equation}\label{stringy_instantons}
\Delta V_{string}= m_{susy}^2M_P^2e^{-2\pi/\alpha_i}\cos(a/F_a-\delta)\,.
\end{equation}
Here $\alpha_i$ stands for the dark sector gauge coupling, or the coupling of any other gauge sector coupled to the light axion, at the compactification scale.   

There are two conditions that need to be satisfied if we want to have quintessence from maximal axion misalignment:
\begin{equation}\label{conditions_stringy_instantons}
(m_{susy}^2M_P^2)^{1/4}e^{-\pi/(2\alpha_i)}\lesssim H_I\,,\,\,\text{ and }\,\,m_{a}^{string}\lesssim H_0\,.
\end{equation}
The first condition tells us that the potential coming from string instantons is \textit{turned off} during inflation so that the axion slow-rolls to the minimum which is determined by the YM hidden sector contribution. The second implies that the current evolution is, also, uniquely determined by the YM generated mass: the stringy mass contribution is \textit{frozen} today and the tachyonic instability of the axion field is not sensitive to this contribution. One easily sees that the first condition typically implies the second automatically, and can be easily achieved (assuming for example $m_{susy}\sim 10^4$ GeV) for gauge couplings at the compactification scale:
\begin{equation}
\alpha_i\lesssim 1/35\,.
\end{equation}
For smaller gauge couplings, the stringy instanton contribution can be safely considered negligible. On the other hand, if the SUSY scale turns out to lie close the Planck scale, then we will need a dark gauge coupling at the compactification scale: $\alpha_i\lesssim 1/45$. We see that the exponential dependence on $\alpha_i$ softens the dependence on the SUSY breaking scale. Either of these possibilities seem therefore plausible to occur in typical compactifications\footnote{Note that we have taken a conservative approach assuming that the typical height of the potential generated by string instantons is $\Lambda_{string}^4\sim M_P^4e^{-S}$ up to SUSY corrections \cite{Arvanitaki:2009fg,Svrcek:2006yi,Hui:2016ltb}. It may happen that in certain compactifications, the potential height is proportional to the string scale $l_s^{-4}\sim M_s^4$. In that case, since we generically consider a low string scale, the corrections of string instantons to the YM axion potential will certainly be smaller, allowing slightly larger values of $\alpha_i$. }. 

Interestingly enough, a fast analysis of the situation seems to disfavor the possibility of coupling the axion that composes maximally misaligned quintessence to the $SU(2)_L$ gauge sector. If we consider a theory with low-energy supersymmetry, the gauge coupling of $SU(2)_L$ is not asymptotically free above the SUSY scale and is always larger than $\alpha_{SU(2)_L}\sim1/35$ unless $m_{susy}\gg O(10^3)$ TeV. This last possibility resembles a pure SM situation, where we have $\alpha_{SU(2)_L}\lesssim 1/45$ only for energies above $E\sim 10^{15}$ GeV, therefore requiring $R^{-1}>10^{15}$ GeV. However, as we have argued before, in the absence of an alternative mechanism, such a large compactification scale is disfavored by the required tuning for the modulus mass, $m_\phi$. Fortunately, there exists the possibility that our axion inherits the coupling to EM at low energies through the coupling to $U(1)_Y$. In the MSSM (SM), $\alpha_Y\lesssim 1/35\, (1/45)$ for any energy scale below $E\sim 10^{11-12}$ GeV and therefore, as long as $R^{-1}\lesssim 10^{11-12}$ GeV, the kind of effects in Eq.(\ref{stringy_instantons}) are expected to be negligible and the conditions in Eq.(\ref{conditions_stringy_instantons}) will be satisfied. In some sense, the requirement of a light axion coupled to photons that realises the maximal-misalignment mechanism shapes the far UV of our theory. Despite the arguments given here are somewhat heuristic, we see that the consistence of the mechanism is non-trivial and limits the couplings that this kind of axions have to different gauge sectors. Of course, being less sensitive to the particular initial conditions, the case of maximally-misaligned ultralight axion DM (see Sec.(\ref{sec:ULA})) is not so sensitive to the size of the gauge couplings.

\section{Discussion}\label{sec:Discussion}
We have considered theories where the inflationary axion potential and the axion potential in vacuum today coincide. This occurs whenever $H_I\lesssim (m_aF_a)^{1/2}$ and therefore requires a low inflationary Hubble scale, generically $H_I\lesssim O(100)$ eV (see Eq.(\ref{relic_small_misalignment})), if we want to avoid problems of overabundance of the many axions in the axiverse. In this case,  when $H_I\ll O(100)$ eV, we have also shown that obtaining a non-negligible abundance of DM requires to shift the minimum of the axion potential that composes the DM between inflation and today. To this end, we have introduced the maximal-misalignment mechanism as a prototype example that, in addition to DM, allows to obtain axion quintessence to describe dark energy with a decay constant well below $M_P$. Here we comment on alternative possibilities to regulate the axion contribution to the DM abundance.

As we briefly mentioned in section \ref{sec_stochastic_axiverse}, a different possibility could be to have an axion mass during inflation  much larger than its current value.  It is well known that scalar fields that are heavy during inflation, $m\geq H_I$, fastly roll towards the minimum of the potential. One could therefore impose that, during inflation, all the dangerous axions in the axiverse are considerably heavier than today. To illustrate this, let us individually study the case of the QCD axion. Let's imagine that the QCD scale during inflation is much larger than today's value:
\begin{equation}
(\Lambda_{QCD}^{inf})^4\gg \Lambda_{QCD}^4\sim (75.5\text{ MeV})^4\,.
\end{equation}
This may occur in string scenarios where all the continuous parameters, including gauge couplings, are given by the expectation value of moduli fields. As argued in \cite{Dvali:1995ce}, since the initial values of the moduli fields are usually far away from their true minima if $H_I$ is large, the inflationary gauge coupling might be larger than today's value allowing QCD to become strong during the inflationary phase. This would result in $\Lambda_{QCD}^{inf}\gg \Lambda_{QCD}$. As explained in \cite{Choi:1996fs}, this scenario is challenged by the fact that the low energy minimum of the axion potential, given by the physical $\bar{\theta}$, includes a number of CPV parameters in addition to the bare $\theta$-term. These include the phase coming from the Yukawa couplings and possibly additional parameters, i.e. phases, in similarity to SUSY scenarios or theories with additional fermions. In scenarios where gauge couplings are determined by the expectation value of a modulus field, it is plausible that these additional sources of CPV also depend on the value of other moduli fields which render their value random in the early Universe. It is therefore unlikely that both vacua, the inflationary and today's vacuum, coincide for all the axions appearing in the axiverse in theories with a large $H_I$. It has been proposed \cite{Banks:1996ea,Co:2018phi} that a theory where CP is an exact symmetry, only broken spontaneously by the VEV of a field, together with a CP invariant inflaton sector, could explain why the inflationary minimum and today's vacuum coincide up to small corrections. However, this situation is challenged by the fact that additional contributions to the axion potential (e.g. string instantons or other YM instantons breaking different axion shift symmetries) may also be enhanced during inflation, due to displaced moduli, in the same way as the QCD contribution.
In general, it is therefore unclear that both vacua will coincide in situations where the axion mass during inflation is much larger than the current axion mass in vacuum. This situation would bring us back to a scenario with a random initial angle. In addition, even if a consistent mechanism is found, many axions in the axiverse may need to be \textit{regulated} in a similar way, by the same mechanism. We conclude that this sort of solutions are very sensitive to the initial conditions of moduli that fix the gauge couplings and other parameters and, despite explicit realisations might be constructed, in general, seems not be the best way to regulate the axiverse cosmology.
\\\\
Of course, there is always the possibility that most  of the axions in the axiverse have a decay constant in the regime: $F_a\ll M_P$. Indeed, this seems to be the case in many of the points in moduli space that lie well inside the stretched K\"ahler cone, where the volume of cycles are at least $O(1)$ in string length units \cite{Demirtas:2018akl} and computational control can be guaranteed. In this case, the associated decay constants are distributed around a mean value that decreases as the number of axions increases. Just as an example, in their dataset (see \cite{Demirtas:2018akl,Mehta:2021pwf} for details), the decay constant follows a log-normal distribution that for $n\sim O(100)$ axions is centered around $\vev{F_{a}}\sim 10^{12-13}$ GeV. The concrete value depends on the position in the K\"ahler cone. Regarding the axion masses, \cite{Demirtas:2018akl,Mehta:2021pwf} confirm that, apart from a peak that looks narrower for points well inside the K\"ahler cone, the distribution is approximately log-flat, with axions populating almost all mass scales from $M_P$ down to $H_0$. If this is the case for our Universe, an axiverse with many axions with masses distributed uniformly and  $\vev{F_{a}}\ll M_{GUT}$, several signals such as BH superradiance might be difficult to be observed due to the effect of axion self-interactions that arises when $F_a\ll M_{GUT}$. Even in this case, due to the log-flat distribution of their mass, several axions may need to be regulated to reduce their contribution to the relic abundance, rendering the class of mechanisms proposed here a compelling possibility to achieve a cosmologically consistent picture of the axiverse.  
\newline\newline
Finally, let us comment on the compatibility of maximally-misaligned quintessence and $N$-essence with the bounds imposed by the string swampland criteria \cite{Vafa:2005ui}. In the last years a lot of attention has been driven to these constraints in the context of theories of inflation and dark energy (see \cite{Palti:2019pca} for a comprehensive review).  
While certain conjectures such as the Weak Gravity Conjecture \cite{ArkaniHamed:2006dz}, the absence of global symmetries \cite{Banks:1988yz} and the distance conjecture have accumulated a certain amount of evidence during the last years, others such as the de Sitter conjecture \cite{Obied:2018sgi} is, in some sense, still in a premature stage. 
It is nevertheless an interesting exercise to ask whether the models of maximally-misaligned quintessence and $N$-essence satisfy these criteria and, at the same time, can describe the current epoch of accelerated expansion.

The swampland distance conjecture states that the distance traversed by scalar fields  is bounded, in  field  space, by $\Delta\Phi\sim O(1)$ in Planck units.  This is easily satisfied in our framework since the inequality: $NF_a^2\ll M_P^2$ is guaranteed in most of the parameter space with successful $N$-essence (see Fig.(\ref{fig_Nessence}))  and always satisfied for maximally-misaligned (single axion) quintessence. Notice that, even in the $N$-essence case, there are no transplanckian scales hidden in any change of basis - there is no direction in the moduli space that is larger than $M_P$. This occurs because, in the context of $N$-essence, there are two sources of \textit{decay constant reduction} with respect to the Planck mass: small offset with respect to the top of the potential (apparent exponential fine-tuning) and existence of $N$ light axions realising the maximal-misalignment mechanism.  The combination of both allows a Pythagorean sum smaller than the Planck scale, $NF_a^2\ll M_P^2$, and in some cases with moderately large number of axions, $N\sim O(10)$, we can achieve $F_a\sim M_{GUT}$. Therefore the $N$-essence scenario is safe from quantum gravity effects and compatible with the WGC.

On the other hand, the refined de Sitter conjecture (RdSC) \cite{Garg:2018reu,Ooguri:2018wrx} states that any scalar potential coupled to gravity has to satisfy:
\begin{equation}
M_P\frac{|\nabla V|}{V}\geq c \,,\,\,\text{ or }\,\,\text{min}\left(\nabla_i\nabla_jV\right)\leq-c^\prime\frac{ V}{M_P^2}\,,
\end{equation}
with $c,c^\prime$ $O(1)$ numbers. It is trivial to see that while the first condition is not satisfied at the top of the potential, the second is easily achieved. The curvature at the top of the axion potential - where we can have the accelerated expansion - has negative curvature and satisfies the RdSC bound provided $F_a\leq M_P$. As noted in \cite{Ooguri:2018wrx}, for any axion potential of the type: $V(a)\propto (1-\cos(a/F_a))$, the WGC ensures the RdSC condition. Finally, we remark here that due to the tachyonic instability of the axion potential in the concave region, it is precisely satisfying the WGC, and therefore the RdSC, what makes difficult to have axion quintessence in the absence of tuning \cite{Agrawal:2018rcg} or a dynamical mechanism like the one proposed here.
\section*{Acknowledgments}
I would like to thank Prateek Agrawal, Junwu Huang, Clara Murgui, Davide Racco and Jos\'e Valle for many  useful discussions and comments on the draft. I also thank Miguel Escudero for helpful discussions. This work is supported by the grants FPA2017-85216-P(MINECO/AEI/FEDER, UE), SEJI/2018/033 (Generalitat Valenciana), the Red Consolider MultiDark FPA2017-90566-REDC and FPU grant FPU16/01907.
%\newpage
\appendix
\section{More on the tachyonic instability and multi-instanton potentials}
Throughout this paper we have assumed a single cosine potential for the axion. In this appendix we study in more detail the tachyonic instability of the axion when its initial condition is $\theta\approx\pi$.  We also show that the single cosine potential offers an approximate but reliable estimation of the time during which the axion behaves as dark energy even if there are higher harmonics. In fact, the single cosine constitutes a conservative estimate.

If the axion potential is given by $V(\theta)=\Lambda^4(1-\cos(\theta))$, then we can expand close to the maximum:
\begin{equation}
V(\theta)=2\Lambda^4-\frac{m^2}{2}(\theta-\pi)^2+...
\end{equation}
Defining the offset with respect to the top of the potential as: $\delta\theta=(\theta-\pi)$, the EOM for for this offset reads:
\begin{equation}
\ddot{\delta\theta}+3H\dot{\delta\theta}-m^2\delta\theta=0\,.
\end{equation}
This equation has a solution (see also \cite{Kaloper:2005aj}):
\begin{equation}
\delta\theta(t)=\delta\theta_ie^{\alpha_{\pm}t}\,,
\end{equation}
with
\begin{equation}
\alpha_{\pm}=\frac{1}{2}\left(-3H\pm\sqrt{9H^2+4m^2}\right)\,.
\end{equation}
Only $\alpha_+\approx m$ has physical meaning.

In general, in a gauge theory that becomes strongly coupled in the IR, there can exist multi-instanton contributions with coefficients comparable to the single instanton contribution. Here we show that, under reasonable circumstances, these contributions only make the potential \textit{flatter} at the top, compared to the curvature at the minimum. Let's now assume that:
\begin{equation}
V(\theta)=\sum_n \Lambda_n^4(1-\cos(n\theta))\,.
\end{equation}
Note also that the higher harmonic corrections are necessarily aligned with the leading term by virtue of the Vafa-Witten theorem \cite{Vafa:1984xg}. Close to the minimum, at $\theta=0$, we have
\begin{equation}
V(\theta)\approx \frac{1}{2F_a^2}\sum_n \Lambda_n^4n^2\theta^2+\text{higher orders}\,,
\end{equation}
meaning that the quantity $m^2=\frac{1}{2F_a^2}\sum_n \Lambda_n^4n^2$ defines the axion mass at the minimum. Although they can be in general comparable, to ensure the finiteness of the sum we need a convergence condition. The less restrictive condition looks like: $\Lambda_n^4n^2>\Lambda_{n+1}^4(n+1)^2$.  This is indeed a necessary but not sufficient condition to ensure the convergence of the serie. We can also study the Taylor serie of the potential at $\theta\approx\pi$, which gives us: 
\begin{equation}
V(\theta)=2\Lambda_{total}^4-\frac{\tilde{m}^2}{2}(\theta-\pi)^2+...
\end{equation}
where the total height reads:
\begin{equation}
2\Lambda_{total}^4=\sum_{n=odd} 2\Lambda_{n}^4\,,
\end{equation}
and $\tilde{m}^2$ corresponds to the curvature at the top, which determines the cosmic evolution once we take into account the effect of the tachyonic instability.
If the axion mass $m^2$ is fixed to a finite value of interest (such as comparable to $H_0$ or any other interesting scale), then the absolute value of the curvature at the top:
\begin{equation}
\tilde{m}^2=\frac{1}{2F_a^2}\sum_n \Lambda_n^4n^2 (-1)^{n+1}\,,
\end{equation}
is always strictly smaller than the curvature at the minimum. The alternate sign, $(-1)^{n+1}$, ensures $\tilde{m}^2<m^2$. Therefore our previous estimate, which made use of the single cosine potential:
\begin{equation}
\delta\theta(t)=\delta\theta_ie^{m t}\,,
\end{equation}
is a good (in fact, conservative) estimate. This justifies that we focus only in potentials of the type: $V(\phi)\propto(1-\cos\phi)$.
\begin{figure}[t]
	\centering
	\includegraphics[width=0.49\textwidth]{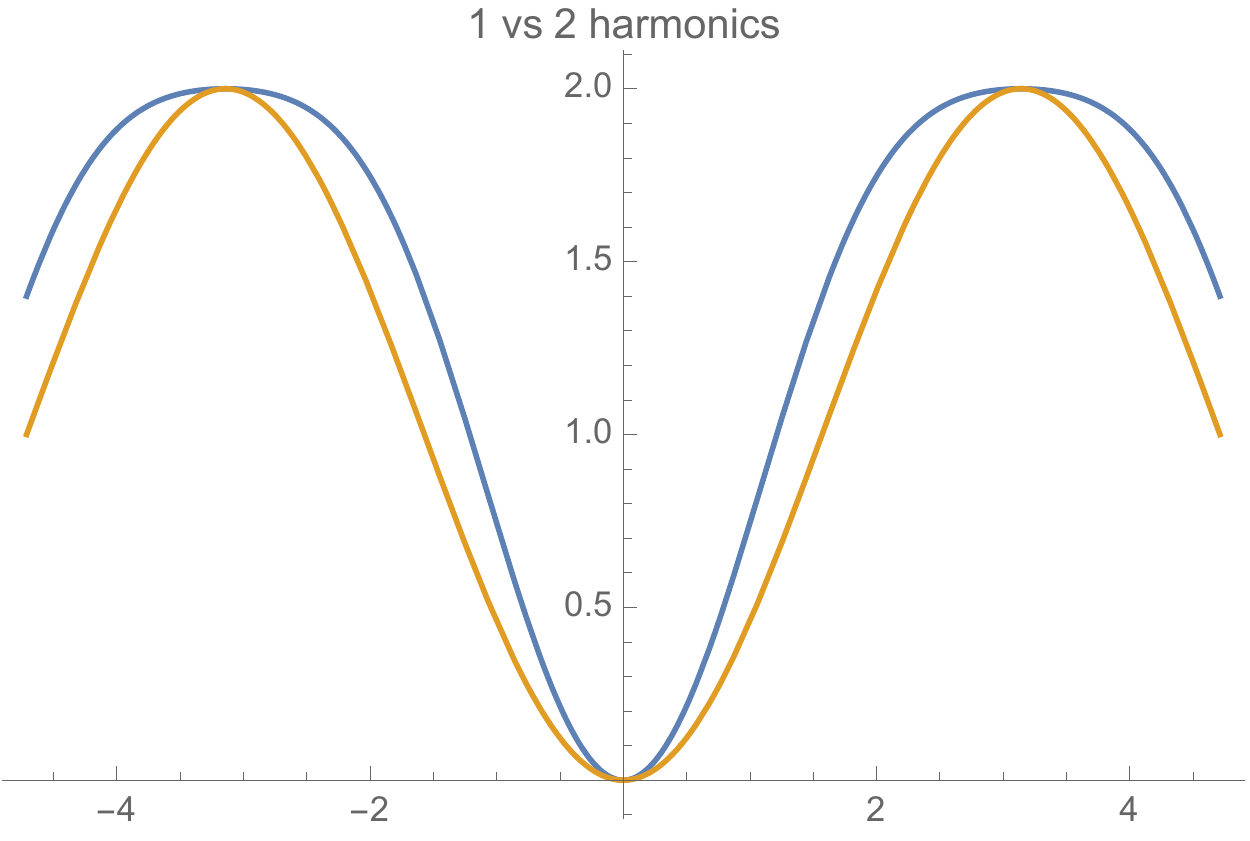}
	\caption{Single cosine (orange) vs 2 harmonic potential (blue). Here we take an example with: $V_{orange}=(1-\cos(\theta))$ and $V_{blue}=(1-\cos(\theta))+0.2 (1-\cos(2\theta))$. For the 2 harmonic potential the curvature at the top is a factor 9 smaller than the curvature at the minimum of the potential.}
	\label{higher_harmonics}
\end{figure}
\section{QCD-like axion potential}
In the presence of light fermions ($m_f\ll\Lambda_D$), as it is the case for QCD, the mixing with mesonic states (light Goldstone modes) gives us a different potential that we can compute with $\chi_{PT}$ techniques:
\begin{equation}
V(a)=-\Lambda^4\sqrt{1-b\sin^2\left(a/2F_a\right)}\,.
\end{equation}
The parameter $b$ depends on the fermion masses and it satisfies $b<1$. This gives a potential that is more \textit{peaked} than the standard cosine potential. We can estimate this effect by taking the ratio of the curvature at the top with respect to the curvature at the minimum:
\begin{equation}
\frac{\tilde{m}^2}{m^2}=\frac{1}{\sqrt{1-b}}\,,
\end{equation}
we see that the curvature at the top is larger, $\tilde{m}^2>m^2$, but not by a large factor. Only in the limit $b\rightarrow 1$, the ratio diverges: it corresponds to the exact isospin limit. For the QCD axion, with $m_d(2 \text{ GeV})\approx 2 m_u (2\text{ GeV})$, this ratio is around $3$ \cite{diCortona:2015ldu}. This kind of potentials also receive higher-harmonic corrections but the mixing part is expected to dominate the behavior of the potential.

%\bibliographystyle{utphys}
%\bibliography{newrefs_axion}

\providecommand{\href}[2]{#2}\begingroup\raggedright\endgroup

\end{document}